\documentclass[tighten,twocolumn,twocolappendix, trackchanges]{aastex701}
\usepackage{amsmath}
\usepackage{array}
\newcommand{\CHARA}{The CHARA Array of Georgia State University, Mount Wilson Observatory, Mount Wilson, CA 91023, USA}

\newcommand{\CHARAGSU}{Center for High Angular Resolution Astronomy and Department  of Physics and Astronomy, Georgia State University, P.O. Box 5060, Atlanta, GA 30302-5060, USA}

\begin{document}
\title{Detection and Astrometry of the Ba–Bb Subsystem in $\alpha$ Piscium: First Dual-Field Interferometry at the CHARA Array}

\shorttitle{Dual-field Interferometry at CHARA: $\alpha$ Psc}

\shortauthors{Anugu et al.}

\author[0000-0002-2208-6541]{Narsireddy Anugu}
\affiliation{\CHARA}
\email{nanugu@gsu.edu}

\author[0000-0002-4313-0169]{Robert Klement}
\affiliation{European Organisation for Astronomical Research in the Southern Hemisphere (ESO) Casilla 19001, Santiago 19, Chile}
\affiliation{Université Côte d’Azur, Observatoire de la Côte d’Azur, CNRS, Boulevard de l’Observatoire, CS 34229, 06304 Nice Cedex 4, France}
\affiliation{\CHARA}
\email{robertklement@gmail.com}

\author[0000-0002-3380-3307]{John D. Monnier}
\affiliation{Department of Astronomy, University of Michigan, Ann Arbor, MI 48109, USA}
\email{monnier@umich.edu}

\author[0000-0001-8537-3583]{Douglas R. Gies}
\affiliation{\CHARAGSU}
\email{dgies@gsu.edu}

\author[0000-0001-5415-9189]{Gail H. Schaefer}
\affiliation{\CHARA}
\email{gschaefer@gsu.edu}

\author[0000-0001-6017-8773]{Stefan Kraus}
\affiliation{Astrophysics Group, Department of Physics \& Astronomy, University of Exeter, Stocker Road, Exeter, EX4 4QL, UK}
\email{S.Kraus@exeter.ac.uk}


\author[0009-0006-9244-3707]{Sebasti\'{a}n Carrazco-Gaxiola} 
\affiliation{\CHARAGSU}
\email{jcarrazcogaxiola1@gsu.edu}

\author[0009-0008-0451-4435]{Akshat S. Chaturvedi} 
\affiliation{\CHARAGSU}
\email{achaturvedi3@gsu.edu}

\author[0009-0006-0225-4444]{Mayra Gutierrez}
\affiliation{Department of Astronomy, University of Michigan, Ann Arbor, MI 48109, USA}
\email{maygut@umich.edu}

\author[0009-0005-8004-2351]{Becky Flores} 
\affiliation{\CHARAGSU}
\email{bflores5@gsu.edu}

\author[0000-0003-3045-5148]{Jeremy Jones} 
\affiliation{\CHARAGSU}
\email{jjones176@gsu.edu}

\author[0009-0005-9132-5779]{Colin Kane} 
\affiliation{\CHARAGSU}
\email{ckane6@gsu.edu}

\author[0000-0002-8955-520X]{Rainer K\"{o}hler}
\affiliation{\CHARA}
\email{rkohler@gsu.edu}

\author[0000-0001-9253-7785]{Karolina Kubiak}
\affiliation{\CHARA}
\email{kkubiak@gsu.edu}

\author{Olli W. Majoinen} 
\affiliation{\CHARA}
\email{omojoinen@gsu.edu}

\author[0000-0003-1038-9702]{Nicholas J. Scott}
\affiliation{\CHARA}
\email{nscott14@gsu.edu}

\author[0009-0006-3626-7585]{Kayvon Sharifi} 
\affiliation{\CHARAGSU}
\email{}

\begin{abstract}
We present the first on-sky demonstration of dual-field interferometry at the
CHARA Array and the first direct resolution of the inner Ba--Bb subsystem in the
bright hierarchical triple $\alpha$~Piscium.
Using $H$-band fringe tracking on component~A with MIRC-X to stabilize $K$-band
science fringes on component~B with MYSTIC, we detected a companion at a
projected separation of 7~mas, confirming a long-suspected
but previously unresolved short-period subsystem within the B component.
The nearly equal $H/K$-band flux ratio indicates that Ba and Bb are near-twin
F-type stars, consistent with the two narrow-lined components seen in
optical spectra of B.
By combining CHARA interferometry with archival VLTI/GRAVITY astrometry and
radial velocities from archival and new spectroscopy (NARVAL and ARCES),
we derive a well-constrained orbit with a period
of $P = 25$~d, eccentricity $e \simeq 0.6$, and inclination $i \simeq 65^\circ$,
yielding precise dynamical masses of $1.668\pm0.033\,M_\odot$ and
$1.646\pm0.029\,M_\odot$.
No additional companion is detected down to $\Delta H \approx 5$ at
separations of 0.2--2~AU.
We also obtained dual-field differential astrometry of the wide A--B pair with a
precision of $\sim0.234$~mas at a separation of $1.85''$, with an error budget
dominated by internal delay-line actuators, fringe-tracking performance and chromatic dispersion.
While the long-period outer orbit is not refined by these measurements, their
agreement with the published astrometric orbit provides an on-sky validation of the
CHARA dual-field mode.
These results establish $\alpha$~Psc as a well-characterized hierarchical system
suitable for future benchmark studies and
demonstrate CHARA's new capability for off-axis interferometry and
sub-mas astrometry on arcsecond-scale binaries.
\end{abstract}


\section{Introduction}

Hierarchical multiple systems provide critical constraints on stellar evolution,
angular-momentum transport, and the dynamical pathways that produce short-period
binaries among intermediate-mass stars \citep{Tokovinin2014ARA&A..52..1,
MoeDiStefano2017ApJS..230..15}.  
Among these, short-period A-star binaries embedded within wider triples often
exhibit conditions---reduced large-scale mixing, intermittent tidal interaction,
and stable radiative envelopes---that can favor the development or persistence
of chemical peculiarities in some components \citep{Preston1974, Abt1995ApJS...99..135A,
Netopil2017}.  
Not all hierarchical systems host chemically peculiar (Ap and Am) stars, but their
architectures provide a useful framework for interpreting rotational, magnetic,
and spectroscopic variability in systems where Ap signatures are present \citep{Mikulasek2011}.

The bright, nearby \citep[$50 \pm 2$ pc;][]{Bailer-Jones2018} visual binary system $\alpha$~Psc (HIP~9487) illustrates both magnetic and metallic-line chemical peculiarity, with component A (HD~12447) classified as an Ap (chemically peculiar 2) star and component B (HD~12446) as an Am (chemically peculiar 1) star \citep{Gray1989ApJS}.
This system is among the longest-monitored visual
binaries, with astrometric measurements dating back to 1779 \citep{Lee1913ApJ,
Mason2001}.  
Component~A exhibits established chemical peculiarities and rapid rotation,
with a 1.5~d magnetic modulation detected in photometry and radio
observations \citep{Wraight2012MNRAS.420..757W, Das2022ApJ...925..125D}.  
Component~B, by contrast, shows very sharp spectral lines and has long been
suspected to host an equal-brightness SB2 subsystem \citep{Abt1985ApJS}, though
no spectroscopic orbit or astrometric confirmation has previously been obtained.

Because the primary goal of this project was to commission and validate the 
new observing mode at the Georgia State University Center for High Angular Resolution Astronomy (CHARA) Array \citep{tenBrummelaar2005}, we selected the bright (H$\sim3.6$~mag) and well separated 
$\alpha$~Psc A--B system for an on-sky demonstration. 
Its wide separation, and almost equal flux double stars,  make it an ideal system for testing fringe tracking, dual-channel
fiber injection, and differential-delay control prior to observations of fainter
science targets. Because a dual-field capability has long been a missing step in CHARA’s evolution, $\alpha$ Psc offered a rare opportunity: a bright, well-understood system in which we could both commission the technique and potentially uncover unresolved structure.

\begin{figure}
\centering
\includegraphics[width=0.49\textwidth]{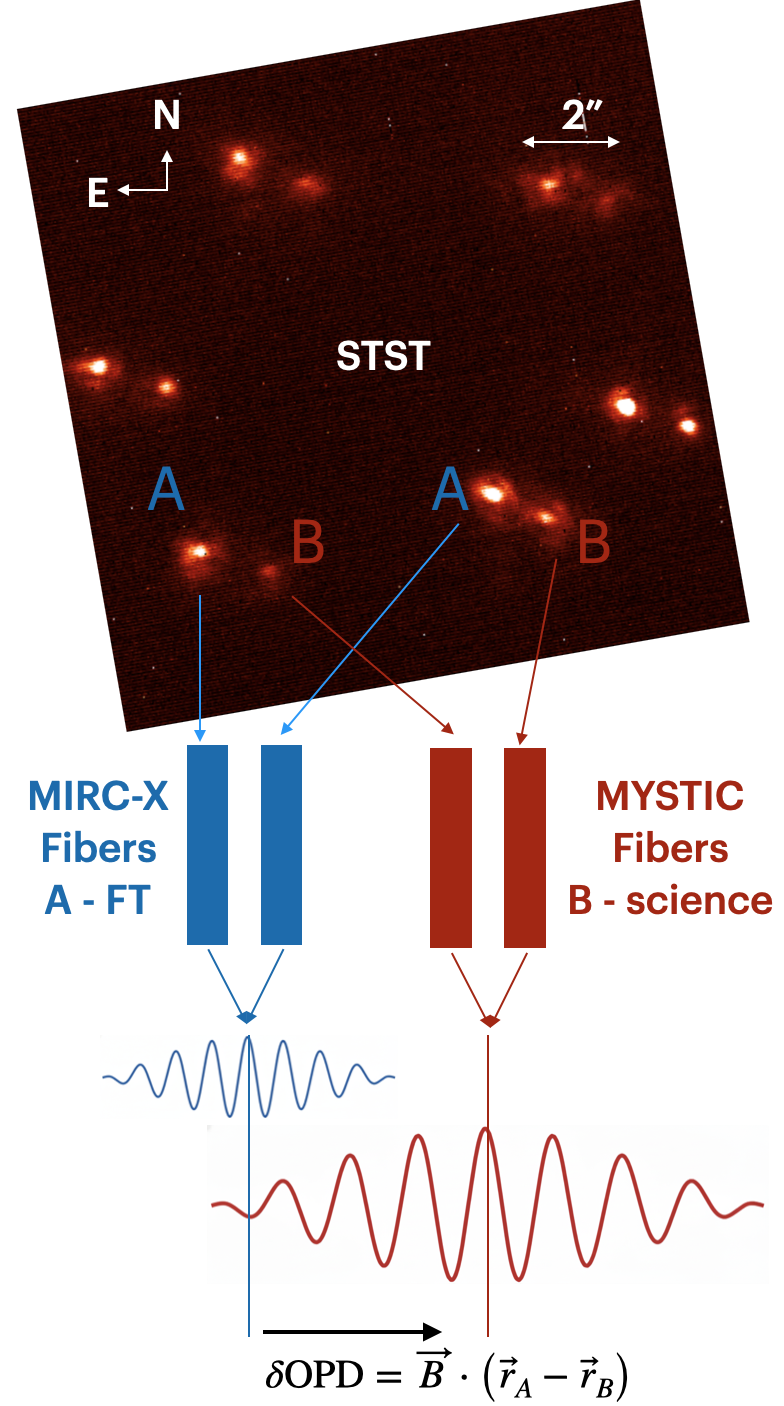}
\caption{Light from the six 1-m CHARA telescopes is relayed to the beam-combiner laboratory and imaged on the STST (Six Telescope Star Tracker) camera with a field of view of $\sim 5^{\prime\prime}$ \citep{Anugu2025SPIE}. 
Using the STST camera coordinates of the $\alpha$~Psc~A--B pair, flux injection into the single-mode fibers is optimized: the brighter A component into the MIRC-X H-band fibers (blue), and the fainter B component into the MYSTIC K-band fibers (red). 
Each fiber pair (two-telescope fringes shown for clarity) produces a corresponding white-light fringe packet, whose relative offset is set by the sky separation of the two stars. Here we define the differential optical path difference (OPD) between the two stars (and thus between the two channels) as $\delta\mathrm{OPD}$.
The measured fringe-packet separation $\delta\mathrm{OPD}$ therefore yields the instantaneous on-sky astrometric offset between the two stars with sub-mas precision.
This dual-field scheme enables simultaneous fringe tracking on the bright component and long-exposure science measurements on the faint component. See detailed optical layout in Figure~\ref{fig:dual_star_optical_layout}  and implementation in Appendix~\ref{app:dualfield}.
}
\label{fig:dualfield_overview}
\end{figure}

In this paper we report dual-field interferometric observations at
CHARA (see Figure~\ref{fig:dualfield_overview}) and use them to (1) directly
resolve the previously unseen Ba--Bb subsystem at a separation of $\sim$7~mas (see Figure~\ref{fig:BaBb_orbit}) and (2) measure the A--B separation via differential delay,
providing an on-sky validation of the new astrometric mode (see Figure~\ref{fig:AB_orbit}).
This newly resolved subsystem clarifies the spectroscopic complexity of the B
component and establishes $\alpha$~Psc as a well-characterized system for studying hierarchical
dynamics in intermediate-mass stellar systems.

The paper is organized as follows.  
Section~\ref{sec:technique} describes the observational setup and data
reduction.   
Section~\ref{sec:BaBb} details the discovery and orbit of the Ba--Bb subsystem
using interferometric and archival spectroscopic data.  
Section~\ref{sec:A} analyzes CHARA and TESS \citep{Ricker2014} observations of component~A,
demonstrating rapid rotation and confirming the absence of a close companion.  Section~\ref{sec:Astrometry_AB} presents the dual-field astrometry of the wide
A--B pair. 
Section~\ref{sec:Discussion} discusses the broader astrophysical implications
and future prospects of CHARA dual-field interferometry.  
Technical details of the dual-field implementation, astrometric error budget and validation (see Appendix~\ref{app:dualfield}), observing log, data reduction, 
orbital fitting,  and TESS analysis appear in the Appendices.
Tables~\ref{tab:system_properties}, and \ref{tab:BaBb_orbit_main}, summarize the $\alpha$ Psc system properties and the Ba--Bb
orbital solutions.

\begin{table}
\begin{center}
\caption{
Basic properties of the $\alpha$~Psc system components.
}
\begin{tabular}{lccc}
\hline\hline
Parameter & A & Ba & Bb \\
\hline
Sp. type$^{a}$ &
kA0hA7SrCr &
\multicolumn{2}{c}{kA2hF2mF2}  \\
$d$ (pc)$^{b}$ &
\multicolumn{3}{c}{$48.17\pm0.34$} \\
$M$ ($M_\odot$)$^{c}$ &
$2.55 \pm 0.11$ &
$1.668\pm0.033$ &
$1.646\pm0.029$ \\
$P_{\rm rot}$ (d)$^{d}$ &
$1.491\pm0.022$ &
--- &
--- \\
$R$ ($R_\odot$)$^{e}$ &
$2.33\pm0.26$ &
$1.76\pm0.41$ &
$1.55\pm0.41$ \\
\hline
\end{tabular}
\end{center}
\label{tab:system_properties}
\tablecomments{
$^{a}$ Spectral types from \citet{Gray1989ApJS}.\\
$^{b}$ Distance derived from our combined astrometric and radial-velocity orbital solution (see Section~\ref{sec:BaBb_orbit}). \\
$^{c}$ Mass of component~A from \citet{Netopil2017}. Other literature estimates for component~A based on position in the HRD and evolutionary tracks include $M=2.65^{+0.21}_{-0.28}\,M_\odot$ \citep{Sikora2019MNRAS.483.2300S} and $M=2.49 \pm 0.05\,M_\odot$ \citep{Shultz2022MNRAS.513.1429S}. Masses of Ba and Bb are derived from our combined orbital solution (see Section~\ref{sec:BaBb_orbit}).\\
$^{d}$ Our TESS analysis measured rotational period of component~A matches with previous studies of magnetic field variability and radio bursts \citep{Borra1980ApJS...42..421B,Sikora2019MNRAS.483.2300S,Das2022ApJ...925..125D}.\\
$^{e}$ Stellar radii derived from limb-darkened angular diameters from our CHARA interferometry, converted to absolute measurements using the distance from the orbital solution (see Section~\ref{app:pmoired}). 
}
\end{table}

\section{Observations and Data Reduction}
\label{sec:technique}

We observed the $\alpha$~Psc A--B pair on UT 2025 August~31 (${\rm MJD} = 60918$) and September~1 (${\rm MJD} = 60919$) with all six CHARA telescopes arranged in a Y-shaped configuration with baseline separations of $\boldsymbol{B}=34$--331~m \citep{tenBrummelaar2005}.  We implemented the dual-field observations leveraging both the Michigan InfraRed Combiner-eXeter \citep[MIRC-X, $1.5$--$1.75~\mu$m,][]{Anugu2020} and Michigan Young Star Imager at CHARA \citep[MYSTIC, $2.0$--$2.4~\mu$m,][]{Setterholm2023} beam combiners. They provided angular resolution of approximately $\lambda/2\boldsymbol{B}$, corresponding to spatial resolution down to $\sim0.5$ and $\sim0.6$~mas in the H and K bands.

We observed the $\alpha$~Psc A--B pair in four complementary dual-field configurations to validate repeatability and verify that the astrometric solution was independent of the choice of science combiner. Figure~\ref{fig:dualfield_overview} shows the optical layout (see also Figure~\ref{fig:dual_star_optical_layout}); full technical details of the dual-field concept and implementation are provided in Appendix~\ref{app:dualfield}.
First, we injected $\alpha$~Psc~A into MIRC-X and B into MYSTIC, acquired data with MIRC-X as the group delay fringe tracking channel and MYSTIC as the science channel, and then reversed the roles so that MYSTIC served as the fringe tracker and MIRC-X as the science combiner.
We then repeated the full sequence after swapping the fiber injections—injecting A into MYSTIC and B into MIRC-X—again taking data in both fringe tracker/science configurations.
This four-configuration strategy provided two independent H-band and two independent K-band astrometric solutions, enabling robust internal consistency checks for the dual-field measurements \ref{tab:astrometry_error_budget}). Astrometry between the A--B pair followed the dual-field delay relation (Eq.~\ref{eq:dualfield_opd}). Full observing details (fiber injection strategy, calibrations, spectral setup) are compiled in Appendix~\ref{app:obs}.

A full description of the reduction, calibration, and model fitting is provided in Appendix~\ref{app:obs}.  We used the public MIRC-X pipeline \citep{Anugu2020}, version 1.5.0 to reduce raw data and derive
15 squared visibilities ($V^2$) and 10 independent closure phases (of 20 measured) for each wavelength band H and K.

\section{Discovery of the Inner Ba--Bb Subsystem and Its Orbit}
\label{sec:BaBb}

\subsection{MIRC-X and MYSTIC data}

The Ba--Bb subsystem was first discovered unexpectedly: during the commissioning run, the science-channel ``waterfall" plot and power spectrum in both MIRC-X and MYSTIC instruments displayed an additional fringe signature (see Figure~\ref{fig:opd_trend}) inconsistent with the wide A--B geometry. A follow-up observation obtained on the subsequent night confirmed the presence of a rapidly moving companion in component B,  exhibiting a change in position angle of approximately $2.2^{\circ}$ over a one-night interval. We thus applied PMOIRED \citep{Merand2022SPIE12183E..1NM} algorithms for companion search to the B-component data (see Appendix~\ref{app:pmoired}). The newly discovered Ba--Bb binary was found to have an angular separation of $\sim7$~mas both bands, and H/K flux ratios compatible with equally bright components. Table~\ref{tab:system_properties} shows the derived properties, and Table~\ref{tab:BaBb_astrometry} lists the measured astrometric positions and flux ratios.

\begin{figure*}
\centering
\includegraphics[width=\textwidth]{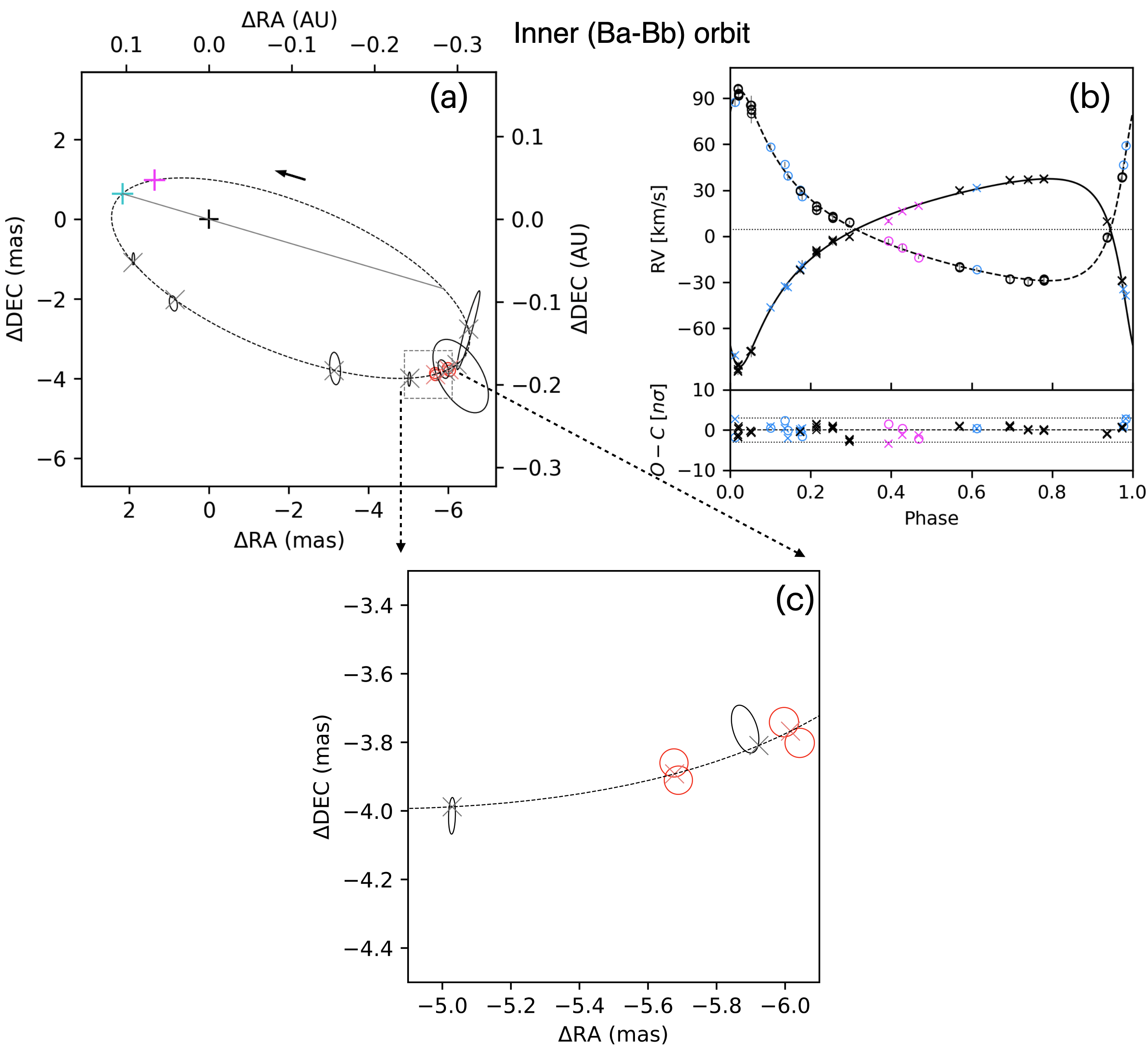}
\caption{
Astrometric and spectroscopic orbit of the inner Ba--Bb subsystem in
$\alpha$~Psc.
\textbf{(a):} Relative astrometry of Ba--Bb, with measurements denoted by 10$\sigma$ error ellipses; CHARA
dual-field measurements obtained in this work are shown in red, while
the archival VLTI/GRAVITY measurements are shown in gray.
The best-fit orbit of the Bb component is represented by the dashed curve, relative to the Ba component (``+'' symbol). The positions on the best-fit orbit for the epochs of our observations are shown as ``x'' symbols. The magenta and cyan ``+'' symbols indicate the position of the periastron and the ascending node, respectively. The line of nodes is shown as a gray line. 
\textbf{(b):} RV curves for Ba and Bb components (solid and dashed curves, respectively), and the corresponding RV measurements (``x'' symbols and open circles). The historical RVs taken from the literature are shown in blue, and our RV measurements from the archival NARVAL and newly obtained ARCES spectra are shown in black and magenta, respectively. The horizontal line corresponds to the systemic velocity. The error bars are too small to be seen for most RVs. The lower panel shows the residuals in units of the measurement errors $\sigma$, with the dotted lines showing $\pm3\sigma$.
\textbf{(c):} Detail of the astrometric orbit showing the measurements as 3$\sigma$ error ellipses.
}
\label{fig:BaBb_orbit}
\end{figure*}

This surprising result prompted us to re-examine all available complementary datasets such as archival spectroscopy and spectropolarimetry, and it became clear why the subsystem had escaped earlier spectroscopic detection: bright visual binaries are
usually excluded from high-precision RV survey
datasets. We also searched for additional archival interferometric data and found a usable dataset from the VLTI/GRAVITY dual-field calibration program (see below).  

\subsection{Archival GRAVITY calibration data}

There are three GRAVITY (GRAVITY Collaboration et al. \citeyear{GRAVITY2017}) single-field snapshots of $\alpha$~Psc available in the ESO archive (program 112.25QQ.002). Two of these observations should have been pointed at the B component according to the data file headers. However, after careful inspection of the B-component data and comparison to the A-component snapshot, we concluded that the two B-component snapshots were inadvertently pointed at the A component, as the data present no binary signature and look almost identical to the A-component snapshot.

We searched for any additional data taken for calibration purposes, and ended up recovering a total of seven usable snapshots that were taken as part of the GRAVITY dual-field calibration program (program 60.A-9801(U)). Each of these snapshots contains data for both the A and B components taken in sequence; first, the A-component is injected into the GRAVITY fringe tracker, while the B-component goes into the GRAVITY science beam combiner, and this is followed with a swap of the components, so that the B-component becomes the fringe-tracking star, and the A-component is the science star. The data reduction, calibration, and analysis of these data is described in Appendix \ref{app:Archival_data}. The PMOIRED analysis of these GRAVITY data resulted in seven additional astrometric points that complement and confirm the findings from CHARA/MIRC-X and MYSTIC, and it also enabled us to obtain the astrometric orbital solution (Section~\ref{sec:BaBb_orbit}, Table~\ref{tab:BaBb_orbit_main}).

\begin{table*}
\begin{center}
\small
\caption{Orbital solutions for the
Ba--Bb subsystem computed with orbfit-lib.  
}
\label{tab:BaBb_orbit_main}
\begin{tabular}{lccc}
\hline\hline
Parameter & Astrometric orbit & Spectroscopic orbit & Combined orbit\\
\hline
Orbital period $P$ (d)               & $24.99925\pm0.00044$ & $24.999444\pm0.000035$ & $24.999428\pm0.000033$ \\
Epoch of periastron $T$ (MJD)        & $59731.09\pm0.13$ & $50930.707\pm0.018$ & $50930.714\pm0.017$ \\
Eccentricity $e$                     & $0.6143\pm0.0051$ & $0.5946\pm0.0031$ &$0.6004\pm0.0018$ \\
Semimajor axis $a$ (mas)             & $4.965\pm0.048$ & --- & $5.175\pm0.018$ \\
Inclination $i$ (deg)                & $60.7\pm1.3$ & --- & $65.20\pm0.40$ \\
Longitude ascending node $\Omega$ (deg) & $75.08\pm0.63^{a}$ & --- & $73.08\pm0.63$\\
Argument of periastron $\omega_{\rm Ba}$ (deg)      & $144.4\pm1.0$ & $139.62\pm0.49$ & $140.14\pm0.42$ \\
RV semi-amplitude $K_1$ (km\,s$^{-1}$)     & --- & $61.02\pm0.34$ & $61.17\pm0.35$ \\
RV semi-amplitude $K_2$ (km\,s$^{-1}$)     & --- & $61.73\pm0.15$ & $61.97\pm0.47$ \\
Systemic velocity $V_{\rm sys}$ (km\,s$^{-1}$) & --- & $4.45\pm0.15$ & $4.52\pm0.37$\\
\hline
Physical semimajor axis $a$ (AU)     & --- & --- & $0.2493\pm0.0020$ \\
Mass ratio $q = M_2/M_1$             & --- &     & $0.9871\pm0.0094$ \\
Total mass $M_{\rm tot}$ ($M_\odot$) & --- & --- & $3.315\pm0.060$ \\
$M_1$ ($M_\odot$)                    & --- & --- & $1.668\pm0.033$ \\
$M_2$ ($M_\odot$)                    & --- & --- & $1.646\pm0.029$ \\
Distance $d$ (pc)                    & --- & --- & $48.17\pm0.34$ \\
\hline
\end{tabular}
\end{center}
\tablecomments{
$^{a}$ There is a $180^{\circ}$ ambiguity in $\Omega$ when fitting purely astrometric orbits. The value listed was chosen to match $\Omega$ when constrained by RV data.
}
\end{table*}

\begin{figure*}
\centering
\includegraphics[width=\textwidth]{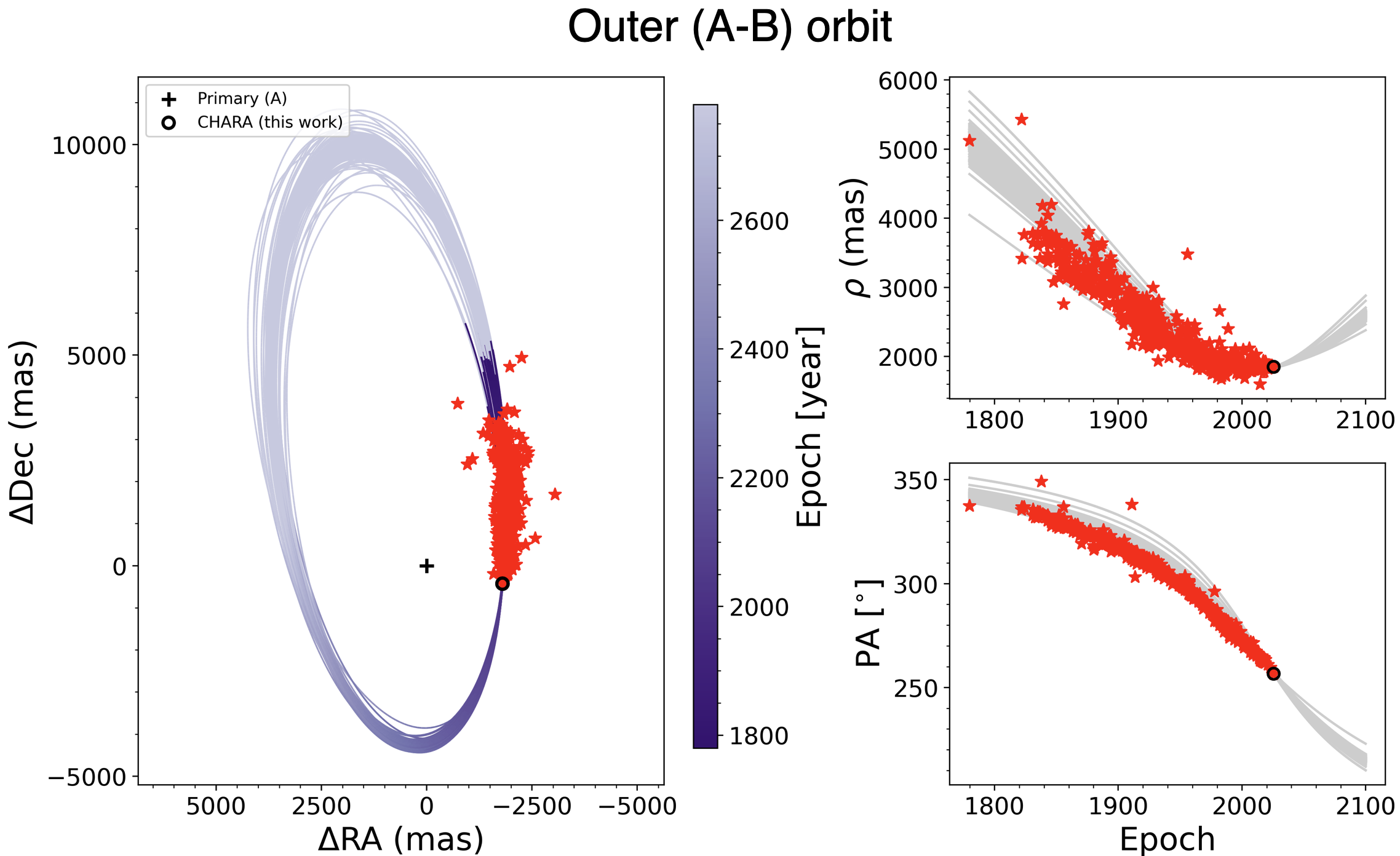}
\caption{Orbit of the $\alpha$~Piscium A--B wide binary.
Gray curves show 200 posterior samples from our MCMC fit to the complete
Washington Double Star astrometric record (1779--2024).
Red points are the historical measurements compiled in ORB6, and the
two black open circles mark the new dual-field astrometric detections
obtained with CHARA in 2025.
North is up and East is to the left.
The CHARA measurements lie exactly along the predicted orbital path at
$\rho \simeq 1.85''$ and ${\rm PA} \simeq 256.6^{\circ}$,
demonstrating that the dual-field mode delivers sub-mas
relative astrometry at arcsecond-level separations.
Including the CHARA points does not materially change the orbital
solution; all elements remain consistent with the published ORB6 orbit at
the $\lesssim 1\%$ level.}
\label{fig:AB_orbit}
\end{figure*}

\subsection{Combined Astrometric + Spectroscopic Orbit}
\label{sec:BaBb_orbit}
Radial velocities (RV) of the Ba and Bb components were measured from 56 archival spectra from the NARVAL spectropolarimetric instrument
at the 2-m Telescope Bernard Lyot,
 and from three new spectra obtained at the ARCES spectrograph mounted on the 3.5-m telescope at the Apache Point Observatory. We also complemented our RV measurements with 10 historical RVs for both components reported by \citet{Abt1980ApJS} and \citet{Abt1985ApJS}. While those earlier studies established the system as double-lined, they did not report a unique orbital period or a full spectroscopic orbit for Ba--Bb. For details on the spectroscopic data and analysis, see Appendix~\ref{app:BaBb_RV}. 

Table~\ref{tab:BaBb_orbit_main} presents the orbital parameters from the purely interferometric/astrometric fit, purely spectroscopic fit, and the final combined astrometric + spectroscopic orbital fit for the Ba--Bb subsystem. The solutions were obtained with the IDL Orbit Fitting Library\footnote{https://www.chara.gsu.edu/analysis-software/orbfit-lib} \citep[orbfit-lib;][]{2006AJ....132.2618S,2016AJ....152..213S}. 
The $25$~d, moderately eccentric ($e\sim0.6$) orbit is now fully constrained,
yielding precise dynamical masses of $1.668\pm0.033\,M_\odot$ and 
$1.646\pm0.029\,M_\odot$ for the two nearly equal-luminosity F-type components, as well as the dynamical parallax of $48.17\pm0.34$~pc.
The Gaia EDR3-based distance to the nearby wide companion HIP~9519 at $404\farcs7$ separation ($48.03 \pm 0.08$~pc; \citealt{Bailer-Jones2021}) matches our Ba--Bb dynamical distance. \citet{Shaya2011ApJS..192....2S} analyzed Hipparcos data and found that HIP~9519 is a physical C-component of the $\alpha$~Psc system (their Table~4, entry \#132 reveals A = HIP~9487, C = HIP~9519). The Gaia distance agreement confirms this association, and the spectral classification of HIP~9519 is G1/2~V \citep{Houk1999}.  
Using Ba--Bb orbital fit distance of $48.17$~pc, the semimajor axis of $5.175\pm0.018$~mas corresponds to $0.2493\pm0.0020$~au. 

Taken together, the wide A--B orbit and the newly constrained Ba--Bb subsystem
establish $\alpha$~Psc as a dynamically stable hierarchical triple in which a
short-period F-type inner binary and a wide tertiary companion supply the
conditions expected for magnetic and chemical peculiarity in
intermediate-mass stars \citep[e.g.,][]{Abt1985ApJS,Auriere2007}.

Details of the orbital sampling, priors, and the incorporation of archival spectroscopic constraints are provided in Appendix~\ref{app:orbit}.

\section{Analysis of $\alpha$~Psc A data}\label{sec:A}

\subsection{MIRC-X + MYSTIC data of $\alpha$~Psc A}
Component~$\alpha$~Psc~A is partially resolved in our H band MIRC-X data, with a uniform-disk (UD)
diameter of $\theta_{\rm UD} \approx 0.45 \pm 0.05$~mas, corresponding to $2.33\pm0.26\,R_{\odot}$ at the distance determined above. No companion is detected brighter than $\Delta H < 5$. To further confirm the result we looked at the archival data from GRAVITY and TESS.  The GRAVITY data for $\alpha$~Psc~A reveal that it is unresolved (compatible with a diameter $<0.5$\,mas at the VLTI baselines) and with no companion signature, confirming CHARA MIRC-X and MYSTIC results.  

\cite{Sikora2019MNRAS.483.2300S} estimate a radius for component~A of $R_\star = 2.66 \pm 0.45\,R_\odot$. For a distance of 48.17~pc, this predicts a limb-darkened angular diameter of $\theta_{\rm LD} \simeq 9.30\,R_\star/d \approx 0.52 \pm 0.09$~mas, consistent with our MIRC-X measurement.

\subsection{TESS Photometric Variability of $\alpha$~Psc A}

For further checking we analyzed TESS data, see Appendix~\ref{app:tess}. The five available TESS sectors (4, 42, 43, 70, 97), spanning seven years, all show a clean, coherent modulation at $P = 1.490975$~d with a strong first harmonic at $P/2 \simeq 0.7455$~d \citep{Borra1980ApJS...42..421B,Sikora2019MNRAS.483.2300S,Das2022ApJ...925..125D}.  
This is consistent with $\alpha$~Psc~A being a magnetic chemically peculiar star, as suggested by prior radio observations \citep{Das2022ApJ...925..125D}.  The phase-folded light curve shows two unequal maxima per cycle and an $\approx$2\% peak-to-peak modulation. 

No significant power is detected at longer periods, and there is no sign of eclipses, ellipsoidal modulation, or other variability that could be associated with the newly resolved Ba--Bb subsystem.


\section{Astrometry and Orbit of $\alpha$~Psc A--B}\label{sec:Astrometry_AB}

Our dual-field differential-delay measurements recover the expected separation
and position angle of the wide $\alpha$~Psc A--B pair, yielding
\[
\rho = 1850.063 \pm 0.234~\mathrm{mas}, \qquad
{\rm PA} = 256.567 \pm 0.009^\circ,
\]
measured on MJD 60918 and 60919, fully consistent with the historical Washington Double Star (WDS) astrometric data (see
Appendix~\ref{app:dual_astrometry_error_budget}). The two nights agree within uncertainty; the expected
day-to-day orbital motion of the A--B pair ($\sim 0.04$~mas~day$^{-1}$) is
well below our sensitivity.
Total mass of the system is $M^{\rm tot}_{\rm AB} = 5.817 \pm 0.165$.
Because the period of the outer orbit (see~Figure~\ref{fig:AB_orbit}) is approximately 3.3~kyr, our two CHARA
epochs sample an extremely small fraction of the orbital phase.  
When incorporated into an updated fit of the A--B orbit
(Appendix~\ref{app:orbit}), the CHARA points fall precisely on the Sixth Catalog of Orbits of Visual Binary Stars (hereafter ORB6\footnote{\url{https://www.astro.gsu.edu/wds/orb6.html}})
trajectory and do not measurably shift any orbital element; all parameters
remain consistent with published values at the $\lesssim 1\%$ level.

Thus, the present A--B astrometry should be regarded primarily as an
on-sky validation of the CHARA dual-field mode---demonstrating
sub-mas relative precision at $1.85''$ separation---rather than as a
revision of the long-period A--B orbit.

A full error budget---including contributions from differential
delay–line repeatability, baseline error, fringe-tracking residuals,
and residual chromatic dispersion---is provided in
Appendix~\ref{app:dual_astrometry_error_budget}.  
These terms combine to yield a total dual-field astrometric precision of
$\sim$0.23~mas. This performance is at the level of the best adaptive optics (AO) based astrometric observations.

Because component B is resolved as a close Ba--Bb pair in the data,
the dual-field differential delay measurement does not trace Ba or Bb
individually, in two short epochs.  Rather, the measured $\delta\mathrm{OPD}$ corresponds to the
flux–weighted photocenter of the B subsystem.  
Given that Ba and Bb have nearly equal $H/K$-band fluxes (within $\sim 2\%$, see~Table~\ref{tab:BaBb_astrometry}),
the photocenter lies within $\lesssim 0.2$~mas of the true barycenter,
 below our current dual-field astrometric precision (see Appendix~\ref{app:impactof_BaBb_A_B}).
Thus, for the purposes of the wide A--B orbit, our measurement effectively
corresponds to the A–B barycentric separation.

\section{Discussion and Conclusions}
\label{sec:Discussion}

\subsection{The Ba--Bb subsystem and system architecture}

Our dual-field CHARA observations directly resolve the long-suspected
Ba--Bb subsystem, establishing that component~B of $\alpha$~Psc is itself
a nearly equal-mass F-type binary with an $H/K$ flux ratio consistent with
unity.  
The presence of two nearly identical components also clarifies why earlier
radial-velocity studies struggled to derive a coherent SB2 orbit at
moderate spectral resolution.

By combining CHARA interferometry with archival VLTI/GRAVITY astrometry
and spectroscopic radial velocities from NARVAL, APO/ARCES, and the literature, we
derive a well-constrained orbit for Ba--Bb, with a period
of $P \simeq 25$~d, eccentricity $e \simeq 0.6$, inclination
$i \simeq 66^{\circ}$, and a physical semimajor axis of
$a = 0.25$~au.
The nearly equal velocity amplitudes and flux ratios imply component
masses of $M_{\rm Ba} = 1.668\pm0.033\,M_\odot$ and
$M_{\rm Bb} = 1.646\pm0.029 \,M_\odot$, yielding a total inner-system mass
of $3.315\pm0.060\,M_\odot$.
These values place Ba--Bb in a region of the mass--period parameter
space that is still underrepresented in high-precision astrometric
samples: short-period ($10$--$30$~d), intermediate-mass F-type binaries
that are bright and nearby enough for sub-mas astrometry remain a
relatively small sample.

Although component~A is a chemically peculiar, rapidly rotating star,
our data do not establish a causal link between the Ba--Bb subsystem and
the Ap characteristics of~A.
Rather, the primary contribution of this work is to provide a complete
and unambiguous system architecture, which is a prerequisite for accurate
atmospheric, evolutionary, and rotational modeling of all components.

The mutual orientation of the two orbital planes can be determined
using the formula from \cite{1981ApJ...246..879F}, and we find that their orientation
is almost perpendicular ($\Phi = 80$ or 119 deg depending on the unknown
factor of $\pm 180$ deg in the longitude of the ascending node $\Omega_{\rm AB}$
for the outer orbit). In such a configuration, the triple can experience
von Zeipel--Lidov--Kozai oscillations that can cause large and cyclic
variations in the inclination and eccentricity of the close binary
\citep{2016ARA&A..54..441N}. One possibility is that such long-term
dynamical interactions may have contributed to the present eccentricity,
although no dynamical modeling is performed here.

\subsection{Dual-field astrometry at CHARA}

As demonstrated in Section~\ref{sec:Astrometry_AB} and
Appendix~\ref{app:dual_astrometry_error_budget}, the wide A--B measurements
primarily serve as an on-sky validation of the new CHARA dual-field mode.
Because the outer orbit has a period of $\sim3.3$~kyr, our two epochs do
not refine the orbital elements.
Nevertheless, the agreement with the ORB6 trajectory demonstrates that
CHARA, in this commissioning experiment, achieves $\sim0.22$~mas relative astrometric precision at
a separation of $1.85''$.
This level of precision is modest compared to the
$\sim25~\mu$as differential astrometry achieved by VLTI/GRAVITY on
sub-arcsecond pairs (GRAVITY Collaboration et al. \citeyear{GRAVITY2018A&A...618L..10G}), but CHARA uniquely
provides access to northern targets and wide ($\lesssim5''$) separations
with long baselines.

In contrast to GRAVITY, which was designed from the outset for narrow-angle
astrometry with a dedicated internal metrology system, the CHARA dual-field
mode is an implementation built on existing infrastructure not originally
optimized for narrow-angle dual-field astrometry.  Our results therefore represent a
first step that demonstrates the feasibility and scientific value of
dual-field operation at CHARA, while highlighting clear paths for future
instrumental upgrades.

As discussed in Appendix~\ref{app:dual_astrometry_error_budget}, targeted
upgrades to the internal differential delay lines (DDL) actuators, use of SPICA-FT 
\citep{Pannetier2022} with $\sim150$~nm phase tracking error and longitudinal wavelength dispersion corrections \citep[LDC,][]{Pannetier2021} are expected to reduce this
floor and enable dual-field astrometry approaching the
$\sim100~\mu$as regime.

\subsection{Summary and outlook}

This work presents the first dual-field interferometric observations
obtained at the CHARA Array and delivers a coherent and unified picture of
the $\alpha$~Psc system.
We (i) directly resolve the B component into a $\sim7$~mas inner binary,
(ii) determine a short-period, eccentric Ba--Bb orbit by combining
interferometry and spectroscopy, (iii) confirm that component~A is a
rapidly rotating, chemically peculiar star with no detected close
companion, and (iv) demonstrate sub-mas dual-field astrometry
on a wide binary.

Beyond the specific case of $\alpha$~Psc, these results mark a critical
step toward extending CHARA’s sensitivity and astrometric reach to faint
companions and hierarchical systems across the northern sky.
A dedicated technical paper is in preparation that will quantify the
end-to-end off-axis faint companion sensitivity performance, coherent integration gains, and
long-exposure limits of the CHARA dual-field architecture, building on
the commissioning results presented here.

\begin{acknowledgments}
We thank J.-B. Le Bouquin, B.~Setterholm, N.~Ibrahim for their contributions to the 
development of the MIRC-X and MYSTIC instruments, and current and previous CHARA team members (V.~Castillo, C.~Farrington,  R.~Ligon,  C.~Lanthermann, H.~McAlister, H.~Renteria, S.~Ridgway, N.~Turner, T.~ten~Brummelaar, N.~Vargas, C.~Woods) for the construction and operational support of the CHARA Array. We are grateful to Dr.\ Rachel Matson of the US Naval Observatory
who provided us with the WDS astrometric data for the A-B pair.
This work is based upon observations obtained with the Georgia State University 
Center for High Angular Resolution Astronomy Array at Mount Wilson Observatory.  
This work also makes use of VLTI/GRAVITY observations obtained under ESO programs 112.25QQ.002 and 60.A-9801(U).
The CHARA Array is supported by the National Science Foundation 
under Grant No. AST-2034336 and AST-2407956. Institutional support has been provided from the GSU College of Arts and Sciences, Office of the Provost, and Office of the Vice President for Research and Economic Development. SK acknowledges funding for MIRC-X from the European Research Council (ERC) under the European Union's Horizon 2020 research and innovation programme (Starting Grant No. 639889 and Consolidated Grant No. 101003096). JDM acknowledges funding for the development of MIRC-X (NASA-XRP NNX16AD43G, NSF-AST 2009489) and MYSTIC (NSF-ATI 1506540, NSF-AST 1909165).
This research has made use of the Washington Double Star Catalog maintained at the U.S. Naval Observatory.
This research has made use of the Jean-Marie Mariotti Center Aspro and SearchCal services. ChatGPT (GPT-5.1) was used to enhance the readability of some parts of the text.
\end{acknowledgments}

\begin{contribution}
NA developed the original research concept, took the CHARA observations, led the analysis, and wrote and submitted the manuscript. RK reduced the VLTI/GRAVITY observations, made the Ba--Bb orbit, and assisted with manuscript revisions. DRG, SCG, ASC, CK, and KS obtained APO RV data. DRG analyzed the archival Ba--Bb radial-velocity data and contributed to manuscript editing. GHS, JDM,  SK and DRG supported the development and operation of CHARA, MIRC-X, and MYSTIC, obtained the associated funding, and contributed to manuscript writing.  All authors reviewed and provided feedback on the manuscript. 
\end{contribution}

\facilities{CHARA (MIRC-X \& MYSTIC), VLTI(GRAVITY), ARC (APO:3.5m/ARCES), TBL (NARVAL), TESS}

\software{PMOIRED \citep{Merand2022SPIE12183E..1NM},
        MIRC-X pipeline \citep{Anugu2020},
        lightkurve \citep{Lightkurve2018},
        JMMC OIFITS and Aspro \citep{Tallon-Bosc2024},
        RVFIT \citep{Iglesias-Marzoa2015},
        orbitize! package \citep{Blunt2020AJ}
          }

\begin{figure*}
\centering
\includegraphics[width=0.8\textwidth]{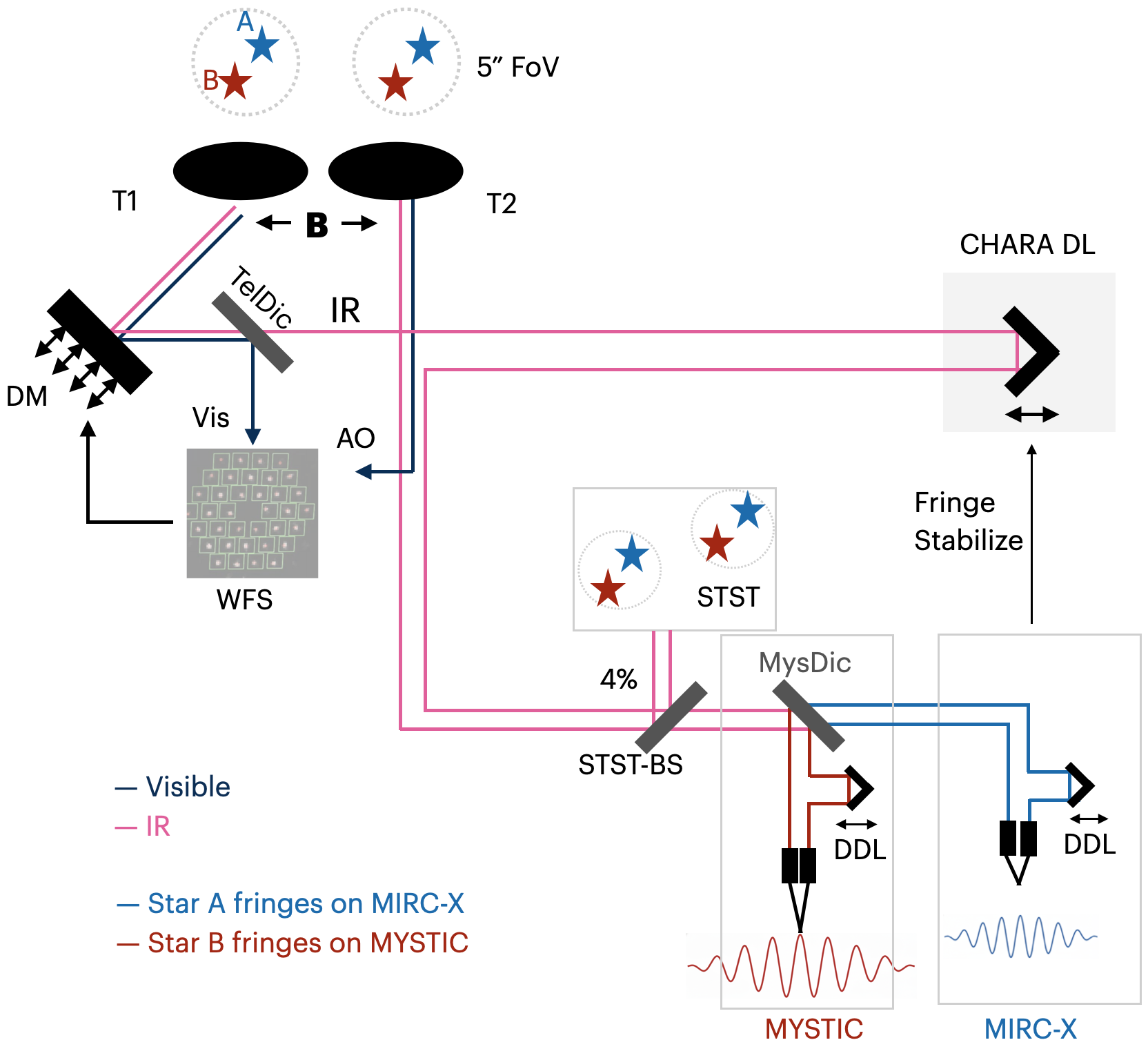}
\caption{Schematic optical layout of the CHARA dual-field mode. For clarity we
show only two telescopes and suppress the second-telescope AO/CHARA delay-line
path. The $\sim5^{\prime\prime}$ field is acquired at the telescopes; visible
light feeds the AO/WFS via the telescope dichroic (TelDic), while the IR beam is
relayed to the beam-combiner lab and main CHARA delay lines. The full field is
imaged by STST, then a small pickoff (STST-BS) sends light to MIRC-X ($H$ band)
and MYSTIC ($K$ band). MIRC-X fringe-tracks on component~A and drives the main
delay lines; MYSTIC records science fringes on component~B (see Figure~\ref{fig:opd_trend}). The instrument DDLs 
apply the additional differential delay $\boldsymbol{B}\cdot\boldsymbol{\theta}$
between A and B, shifting the science beam relative to the fringe-tracking path.
The fiber injection is
set from the STST astrometric positions. Unlike VLTI/GRAVITY, CHARA does not
have an internal metrology system; differential delays are inferred from DDL
telemetry and A-B/B-A swap tests. This schematic illustrates the functional
architecture rather than the full optical complexity of the six-telescope CHARA
beam train.}
\label{fig:dual_star_optical_layout}
\end{figure*}

\begin{figure*}
\centering
\includegraphics[width=\textwidth]{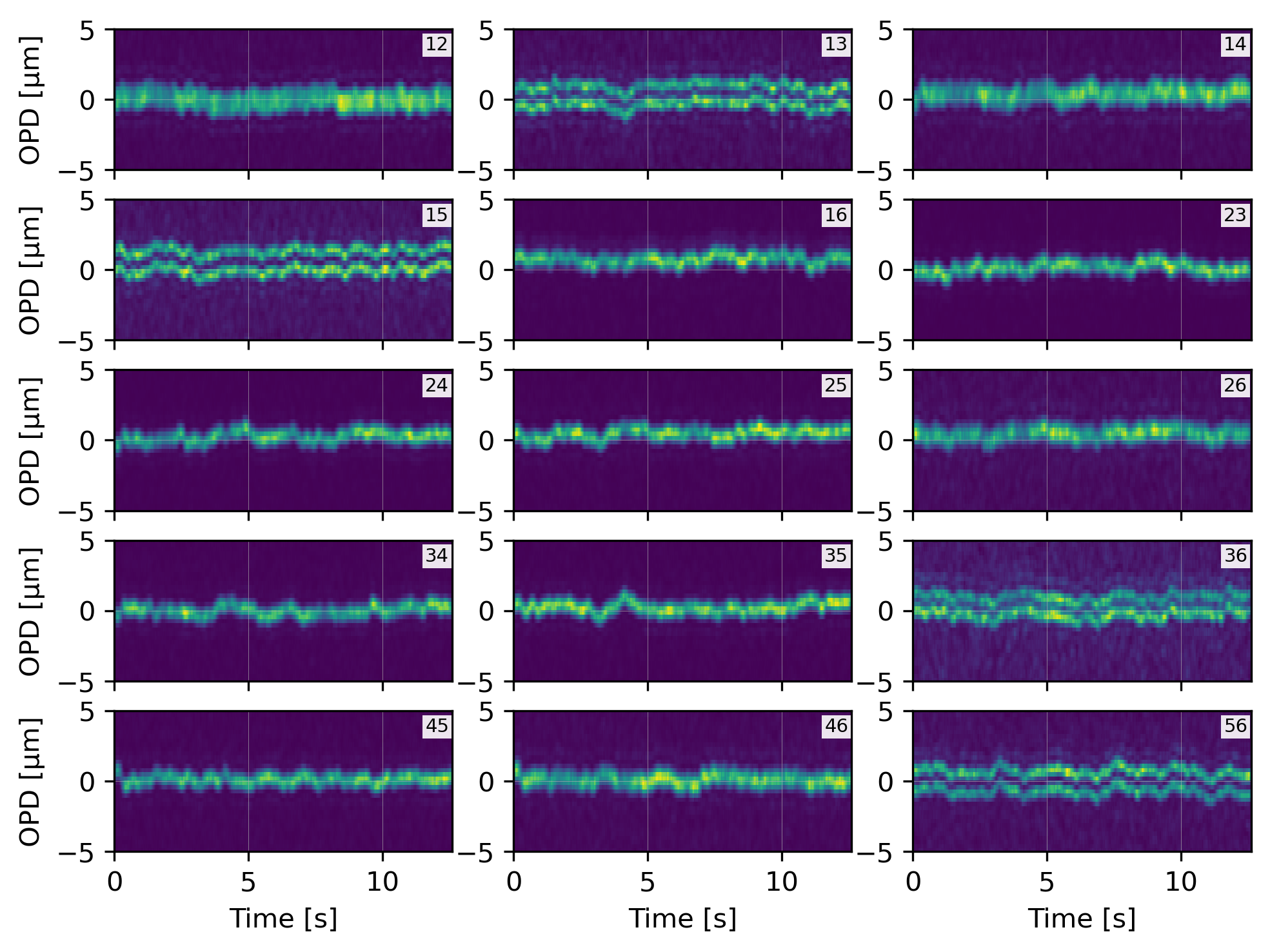}
\caption{Waterfall plots of MYSTIC science-channel OPD residuals during
dual-field observations of $\alpha$~Psc~B (${\rm MJD} = 60918$). The waterfall plot shows the amplitude and position of the Fourier transform of a scan through the interferogram over time. Stable fringe
packets are visible on all 15 baselines despite fringe-tracking residuals,
demonstrating successful phase referencing from component~A and enabling direct
detection of the Ba--Bb subsystem. Each subpanel corresponds to one of the 15
baselines, labeled by the beam pair (e.g., 12 = Telescope 1--2) for the beam
order E1--W2--W1--S2--S1--E2. Time increases horizontally and optical path difference (OPD) is plotted
vertically; brighter lines in each waterfall window indicate higher fringe contrast and successful
fringe detection. The brighter double lines in some of the large baseline waterfall windows are an indication of BaBb binary is resolved.  The coherent fringe integration time for this diagnostic is 10~ms. }
\label{fig:opd_trend}
\end{figure*}

\appendix

This Appendix provides the technical details that complement the results presented in the main paper. 
Appendix~\ref{app:dualfield} describes the dual-field implementation and error budget, Appendix~\ref{app:obs} summarizes the observing setup and data reduction, Appendix~\ref{app:pmoired} details of the interferometric data PMOIRED companion fitting, Appendix~\ref{app:Archival_data} compiles the archival interferometry and spectroscopy, Appendix~\ref{app:orbit} describes the orbital fitting methods, and Appendix~\ref{app:tess} presents the sector-by-sector TESS photometric analysis.  
These materials are provided for transparency and reproducibility, while the primary scientific conclusions---the direct detection of the Ba--Bb pair and the measured day-scale orbital motion---are developed in the main text.

\section{Dual-field Implementation and Astrometric Error Budget}
\label{app:dualfield}

\subsection{Motivation and limitations of single-field operation}

Prior to this work, all CHARA beam combiners---including MIRC-X and
MYSTIC---operated in single-field, on-axis mode.  
In this configuration a single object supplies the flux for tip--tilt
stabilization, adaptive optics correction, fringe tracking, and science
beam combination.  This architecture is efficient for isolated targets,
but its performance degrades for binaries or faint companions located
within a few arcseconds of a bright primary.

For separations $\lesssim 4^{\prime\prime}$, the tip--tilt loop becomes
challenging for the fainter star: in variable seeing individual telescopes
can jump to the brighter component while others remain centered on the
fainter one, preventing fringe acquisition entirely.  
Even when the loop remains nominally stable, the single-field optical
train admits flux from both stars, contaminating the science fringes
with coherent leakage from the nearby star and biasing visibilities and
closure phases.  In practice this makes reliable interferometry on the
faint component of a close pair difficult or impossible.

Single-field operation also limits fringe sensitivity.  
Because the same star must be both tracked and coherently integrated, the
exposure time is restricted by the atmospheric coherence time
($\tau_0 \approx 20$--$40$~ms in $H$ band) and AO corrected coherence diameter
($d_0 \approx 40$--$60$~cm in the near-IR).  
These parameters define an effective coherence volume
($\propto d_0^2\tau_0$): once piston excursions exceed the coherence
length, visibility contrast drops rapidly and the interferometric signal
falls below detection thresholds \citep{Eisenhauer2023}.  
This confines CHARA sensitivity to relatively bright targets
($K \approx 8$--$9$~mag) and leaves many modern science cases---long
coherent integrations on faint companions or substellar objects---out of
reach in single-field mode.

These limitations motivated the development of a dual-field mode at
CHARA, leveraging the existing MIRC-X, MYSTIC, and STST infrastructure
with only modest new observing interfaces.

\subsection{Dual-field interferometry at CHARA}

Dual-field off-axis interferometry addresses both classes of
limitations. It separates fringe tracking and science acquisition
between two stars in the field while enabling narrow-angle differential
astrometry.  
The concept was originally outlined by \citet{Shao1992} and has been
demonstrated at PTI \citep{Colavita1999}, Keck/ASTRA
\citep{Woillez2014}, and most recently at VLTI/GRAVITY, where dual-beam
phase referencing delivers minute-long coherent integrations and
microarcsecond-level astrometry (GRAVITY Collaboration et al. \citeyear{GRAVITY2017}).

Figure~\ref{fig:dual_star_optical_layout} shows that our CHARA
implementation follows the same principle while exploiting the
six-telescope architecture, the existing MIRC-X and MYSTIC beam
combiners, and the STST acquisition camera.  
The internal DDLs in MIRC-X and MYSTIC provide the additional optical
path required to observe two stars separated by
up to $\sim5^{\prime\prime}$ on baselines as long as 331~m.  
CHARA is now the only operational six-beam,
dual-field near-infrared interferometer, and the only facility offering
$H$-band fringe tracking with $K$-band science (and the reverse
configuration).

\subsection{First demonstration on the bright $\alpha$~Psc A--B pair}

Because this commissioning run was designed primarily to demonstrate the
feasibility of dual-field operation (rather than to reach the faintest
possible targets), we intentionally selected the bright ($H\simeq3.6$~mag) and
well-characterized $\alpha$~Psc A--B system  as the
initial on-sky test.
The A--B are within the CHARA dual-field
field of view ($\sim5^{\prime\prime}$; \citealt{Anugu2025SPIE}), while its
separation is larger than the single-fiber core field of view
($\lambda/\boldsymbol{B}_{\rm max} \lesssim 450$~mas in $H$ and $K$). 
On such a high-flux target the MYSTIC science channel saturates before
atmospheric decorrelation becomes limiting, so the effective coherence
time is identical with or without phase referencing.  
For this demonstration, the key performance metrics are:
(i) baseline length knowledge,
(ii) the residual piston delivered by the MIRC-X fringe tracker,
(iii) the stability and repeatability of the DDLs, and
(iv) the relative wavelength calibration and chromatic dispersion between MIRC-X and MYSTIC.
These elements feed directly into the differential-astrometric error
budget presented in Appendix~\ref{app:dual_astrometry_error_budget}.

The optical layout is summarized in
Figure~\ref{fig:dualfield_overview}.  
Light from $\alpha$ Psc A and B is injected into independent different single-mode
fibers:

\begin{itemize}
    \item $\alpha$~Psc~A is injected into MIRC-X ($H$ band) and used as
    the real-time group-delay fringe-tracking reference.  The internal diagnostic
    streams indicate a residual piston of $400$~nm~RMS on
    20~ms timescales.
    \item $\alpha$~Psc~B is injected into MYSTIC ($K$ band), receiving stabilized fringes suitable for coherent averaging and model
    fitting of the Ba--Bb subsystem.
\end{itemize}

Fiber injection for both channels is initialized from STST astrometric
positions and refined by on-sky fiber-map scans. We augmented the existing
fiber-flux optimization/explorer GUI with engineering controls to support
dual-field acquisition and verification.

For each projected baseline $\boldsymbol{B}$ the ideal differential OPD
for two stars with separation vector
$\boldsymbol{\theta}=\boldsymbol{r}_A-\boldsymbol{r}_B$ is
\begin{equation}
    \delta\mathrm{OPD} = \boldsymbol{B}\cdot\boldsymbol{\theta}.
    \label{eq:dualfield_opd}
\end{equation}
Fringe tracking on the reference star is achieved by moving the main CHARA
delay lines \citep{Anugu2026arXiv260216009A}, which are common to both beam combiners. The off-axis science
channel requires an additional differential delay equal to
$\boldsymbol{B}\cdot\boldsymbol{\theta}$; this is provided by the
instrument DDLs, which shift the science beam relative to the
fringe-tracking path until the science fringes are centered (see Figure~\ref{fig:dual_star_optical_layout}). We use these DDL encoder positions $\delta\mathrm{OPD}_{\rm mech}$ saved in the fits data headers and server logs to compute A-B astrometry.

In our instrument setup, the dominant uncertainty arises from the
repeatability uncertainty of the Zaber (T-LA series, 28~mm range)-driven DDL stages, together with fringe-tracking
residuals and residual chromatic dispersion. Although individual DDL moves can show
repeatability error up to the $\pm4~\mu$m level, scan-averaging reduces the DDL repeatability
term to $\sim 0.15~\mu$m. Combined in quadrature with the fringe-tracking
residuals ($\sim 0.25~\mu$m) and residual chromatic dispersion
($\sim 0.2~\mu$m), the effective OPD noise is $\sim 0.36~\mu$m, corresponding
to an astrometric floor of $\sim 227~\mu$as at $1.85''$ separation. The full
error budget is quantified below and summarized in Table~\ref{tab:astrometry_error_budget}.

\subsection{Differential-astrometry formalism and error budget}
\label{app:dual_astrometry_error_budget}

Following the standard narrow-angle interferometry formalism
\citep[e.g.,][]{Shao1992,Colavita1999,Woillez2014}, each projected CHARA
baseline $\boldsymbol{B}_i$ (15 unique baselines for six telescopes)
obeys
\begin{equation}\label{Eq:opd_fit}
   \delta\mathrm{OPD}_i = \boldsymbol{B}_i\cdot\boldsymbol{\theta} + \delta\mathrm{OPD}_{{\rm disp},i} + \delta\mathrm{OPD}_{{\rm inst},i},
\end{equation}
where $\boldsymbol{\theta}$ is the sky-plane separation vector (radians), $\delta\mathrm{OPD}_{{\rm disp},i}$ captures longitudinal dispersion \citep[group-delay offsets between the fringe-tracking and science bands driven by air-path dispersion and effective $H$-$K$ wavelength differences, ][]{Pannetier2021},
and $\delta\mathrm{OPD}_{{\rm inst},i}$ includes residual band-to-band internal offsets.
 We measure five independent OPDs. W2 was used as reference. We measure OPD offsets E1W2, W1W2, S2W2, S1W2 and E2W2.

Throughout this paper, $\delta\mathrm{OPD}$ denotes a differential optical
path delay; subscripts ``meas'', ``disp'', ``inst'', and ``geo'' refer to the
measured delay, chromatic dispersion, residual instrumental offsets, and the
geometric delay computed from the $(u,v)$ coverage, respectively. We use
``mech'' for the mechanical DDL stroke and ``A--B''/``B--A'' to label the
two swap configurations.

\begin{deluxetable}{c l c c c c c}
\label{tab:AB_astrometry_summary}
\caption{Summary of $\alpha$~Psc A--B astrometry from MIRC-X and MYSTIC. }
\tablehead{
\colhead{MJD} &
\colhead{Config} &
\colhead{$\rho$} &
\colhead{PA} &
\colhead{$\sigma_\rho$} &
\colhead{$\sigma_{\rm PA}$} \\
\colhead{} &
\colhead{} &
\colhead{(mas)} &
\colhead{(deg)} &
\colhead{(mas)} &
\colhead{(deg)}
}
\startdata
60918.414 & A-B & 1850.158 & 256.585 & 0.268 & 0.006 \\
60918.427 & B-A & 1850.004 & 256.622 & 0.283 & 0.010 \\
60918.432 & A-B & 1849.671 & 256.607 & 0.160 & 0.015 \\
60918.469 & A-B & 1850.200 & 256.587 & 0.206 & 0.007 \\
60919.372 & B-A & 1850.283 & 256.436 & 0.254 & 0.008 \\
\enddata
\tablecomments{A--B/B--A are defined by the sign of the differential optical path delay: A--B corresponds to $-\delta\mathrm{OPD}$ and B--A to $+\delta\mathrm{OPD}$. We report $\rho$ and PA for the same sky vector $\boldsymbol{\theta}=\boldsymbol{r}_A-\boldsymbol{r}_B$ (A relative to B; PA east of north) in all rows; therefore A--B and B--A entries are expected to agree within uncertainties and are not offset by $180^\circ$.
}
\end{deluxetable}

\begin{table}
\small
\caption{Representative differential-astrometric error budget for a
$1.85''$ binary on the 331~m CHARA baseline.
}
\begin{tabular}{lr}
\hline\hline
Error source & Budget ($\mu$as) \\
\hline
\parbox[l]{0.62\columnwidth}{Baseline length ($\Delta|\boldsymbol{B}| = 10$ mm) }       & $\sim 56$  \\
\parbox[l]{0.62\columnwidth}{DDL scan average error ($\Delta\delta\mathrm{OPD}=0.15~\mu$m)}  & $\sim 93$ \\
Fringe tracking residual  ($0.25~\mu$m)             & $\sim 156$ \\
\parbox[l]{0.62\columnwidth}{Non-common-path OPDs between FT and SC ($H$-$K$ dispersion and drifts, total $0.2~\mu$m)} & $\sim 125$ \\
\hline
Quadrature total                                         & $\sim 227$ \\
\hline
\parbox[l]{0.62\columnwidth}{A-B/B-A swap consistency check (systematic bound; not in quadrature)} & $\lesssim 175$ \\
\hline
\end{tabular}
\label{tab:astrometry_error_budget}
\tablecomments{
Geometric terms scale as
$\delta\theta_{|\boldsymbol{B}|} = \theta(\Delta|\boldsymbol{B}|/|\boldsymbol{B}|)$,
while OPD-related terms scale as
$\delta\theta_{\rm OPD} = \Delta(\delta\mathrm{OPD})/|\boldsymbol{B}|$.
}
\end{table}

Before fitting $\boldsymbol{\theta}$ in Eq.~\ref{Eq:opd_fit}, we clean the OPDs in three steps: (i) remove the on‑axis/off‑axis zero‑point using DDL differencing in Section~\ref{app:off-zero}, (ii) apply the geometric time‑matching step described in Section~\ref{app:geometric}, and (iii) remove the static H–K dispersion via the A--B/B--A swap in Section~\ref{app:swaptest}. The remaining baseline‑dependent offsets are treated as residual chromatic/non‑common‑path terms.

\subsubsection{Systematic-Offset Removal via On-axis–Off-axis DDL Differencing}\label{app:off-zero}

We derive the inter-instrument OPD from the science channel by
comparing MYSTIC DDL positions between two configurations while MIRC-X
continues to fringe track the A component.  First, we place both
instruments on-axis on A and record the mechanical DDL positions (MIRC-X
for fringe tracking, MYSTIC for science).  Next, we keep MIRC-X on A but
move MYSTIC off-axis to B and record the new DDL positions.  The
mechanical differential delay is then
$\Delta\delta\mathrm{OPD}^{\rm mech} =
\delta\mathrm{OPD}^{\rm mech}_{\rm offaxis} -
\delta\mathrm{OPD}^{\rm mech}_{\rm onaxis}$, and the measured optical delay
is $\delta\mathrm{OPD}^{\rm meas} = 2\,\Delta\delta\mathrm{OPD}^{\rm mech}$.
This sequence removes the fixed internal zero-point common to both
pointings and isolates the sky-dependent and chromatic terms.  It does
not remove field-dependent or time-variable aberrations (beam walk and
non-common-path differences between A-B/B-A or between MIRC-X and MYSTIC), so
those residuals are treated as noise terms in the OPD budget.

\subsubsection{Geometric OPD time matching}\label{app:geometric}
As the projected baselines $\boldsymbol{B}_i(t)$ evolve with Earth
rotation, $\delta\mathrm{OPD}_i(t)$ changes. We therefore compute the
geometric model
$\delta\mathrm{OPD}_{\rm geo}(t)=\boldsymbol{B}(t)\cdot(\boldsymbol{r}_A-\boldsymbol{r}_B)$
and time‑match the OPD measurements to the same model timestamps before
comparing the A--B and B--A sequences for dispersion (see~Figure~\ref{fig:ddl_timeseries_babb_2025Aug31}).

\subsubsection{A-B/B-A Swap as a Dispersion Mitigation}\label{app:swaptest}
For H--K bands, $\delta\mathrm{OPD}_{\rm disp}$ is expected to reach up to $57~\mu$m group delay for CHARA air-path differences \citep[from Figure~6 of][]{Pannetier2021}. We did not use the LDC hardware in this experiment; instead,
we mitigate the dispersive term by swapping which component is used for
fringe tracking and which is observed in the science channel (A in
MIRC-X, B in MYSTIC, then the reverse), which flips the sign of
$\delta\mathrm{OPD}_{\rm disp}$. Any residual chromatic OPD is folded
into the dispersion term in Table~\ref{tab:astrometry_error_budget}
(cf.\ the GRAVITY dispersion treatment in \citealt{Lacour2014A&A...567A..75L}).
GRAVITY explicitly tracks metrology--starlight alignment and
metrology-wavelength dispersion; CHARA has no internal metrology, so
those terms are not applicable here and are absorbed into the empirical
OPD-noise budget.

In A-B, configuration (MIRC-X on A, MYSTIC on B) the measured delay is
\begin{equation}
    \delta\mathrm{OPD}^{\rm meas}_{A-B} =
    \boldsymbol{B}\cdot\boldsymbol{\theta}
    + \delta\mathrm{OPD}_{\rm disp}
    + \delta\mathrm{OPD}_{\rm inst}
\end{equation}
where $\delta\mathrm{OPD}_{\rm disp}$ is the longitudinal H--K
dispersion (air/glass group-delay mismatch between the fringe-tracking
and science bands) and $\delta\mathrm{OPD}_{\rm inst}$ absorbs residual
instrumental terms.  After swapping, B-A, the stars (MIRC-X on B, MYSTIC on A)
we measure
\begin{equation}
    \delta\mathrm{OPD}^{\rm meas}_{B-A} =
    -\boldsymbol{B}\cdot\boldsymbol{\theta}
    - \delta\mathrm{OPD}_{\rm disp}
    + \delta\mathrm{OPD}_{\rm inst}
\end{equation}
so both the geometric term and the dispersion term reverse sign.  
The A-B/B-A swap cancels the static component of the H--K
dispersion, but time variability in the air path and effective
wavelengths between the two sequences leaves a small residual, which we
include as a dispersion-noise term in the astrometric error budget.
For example, for the ${\rm MJD} = 60918$ A--B sequence the dispersion offsets are
E1W2: 35.55~$\mu$m, S2W2: 21.10~$\mu$m, S1W2: 27.38~$\mu$m, and
E2W2: 22.59~$\mu$m; these offsets are removed
before astrometric fitting.

\begin{figure}
\centering
\includegraphics[width=0.49\textwidth]{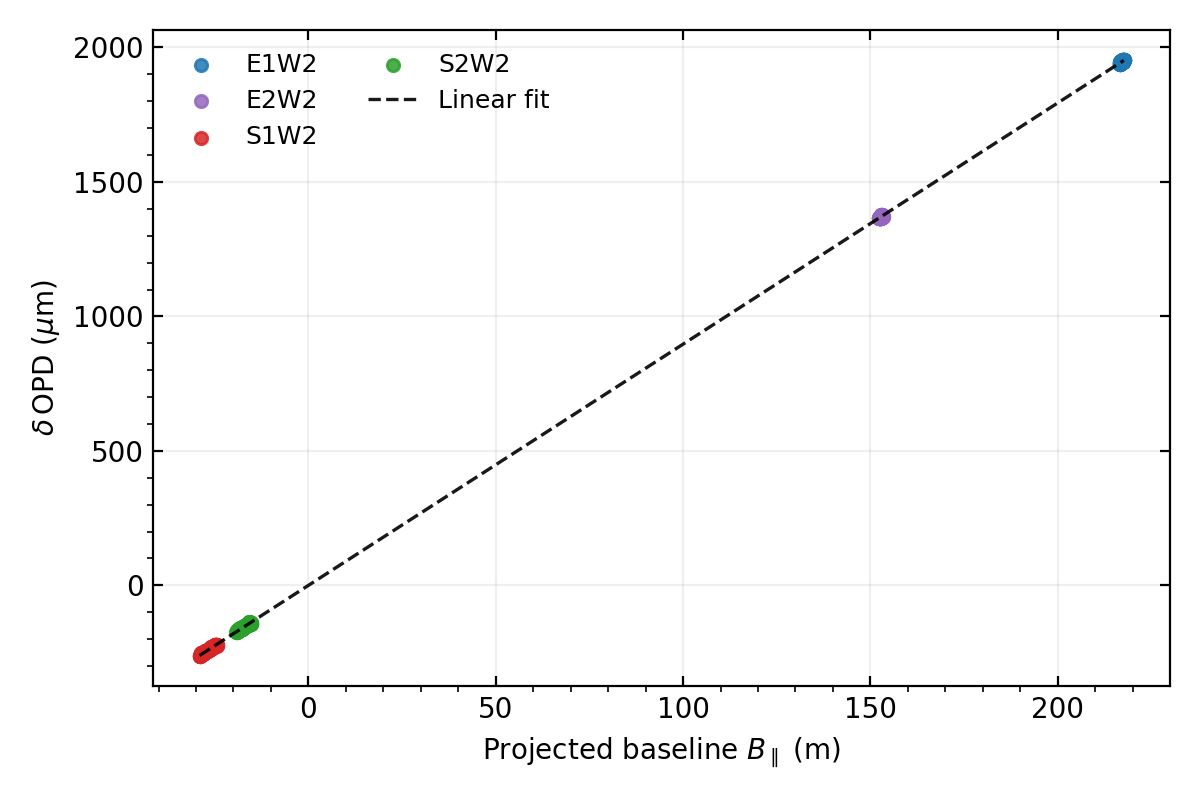}
\caption{Dispersion-corrected $\delta\mathrm{OPD}^{\rm meas}$ plotted against the
projected baseline component $B_\parallel$ for the ${\rm MJD} = 60918$ A--B
sequence. $B_\parallel$ is computed from the $(u,v)$ coordinates via
$B_\parallel=(u\Delta\alpha+v\Delta\delta)/|\boldsymbol{\theta}|$, where
$\Delta\alpha$ and $\Delta\delta$ are the east and north components of
$\boldsymbol{\theta}=\boldsymbol{r}_A-\boldsymbol{r}_B$. Each
color denotes a baseline (W2 reference), and the black line is the
best-fitting linear relation. For a fixed separation vector,
$\delta\mathrm{OPD}$ scales linearly with $B_\parallel$ (the component of
the projected baseline along the A--B separation; using $|\boldsymbol{B}|$
would not yield a linear relation), so the alignment
of the multi-baseline points provides a direct consistency check on the
OPD sign conventions and time matching. The residual scatter about the
fit reflects the $\sim$0.37~$\mu$m effective OPD noise used in the
bootstrap uncertainty estimate.}
\label{fig:opd_vs_b}
\end{figure}

Figure~\ref{fig:opd_vs_b} provides a complementary diagnostic by
plotting the dispersion-corrected OPDs against the projected baseline
component $B_\parallel$ for the same scans; the linear relation expected
from $\delta\mathrm{OPD}=\boldsymbol{B}\cdot\boldsymbol{\theta}$ offers a
compact visual check on sign conventions, baseline geometry, and
residual scatter.

\subsubsection{Astrometric OPD Fit and Bootstrap}
We form $\delta\mathrm{OPD}^{\rm meas}$ from scan-averaged DDL positions,
so the DDL repeatability term is treated as the RMS scatter about the
mean (zero-mean noise) rather than as a systematic offset.  Likewise, the
combined fringe-tracking residuals from the fringe tracker and science
combiner are characterized by the RMS of the detrended piston time
series (total $\sim 0.25~\mu$m; the 10--20~nm RMS contribution from the
main CHARA delay lines is included here, see \citet{Anugu2026arXiv260216009A}) and propagated as a noise term
in the OPD uncertainty (see Figure~\ref{fig:opd_trend}).

\begin{figure*}
\centering
\includegraphics[width=\textwidth]{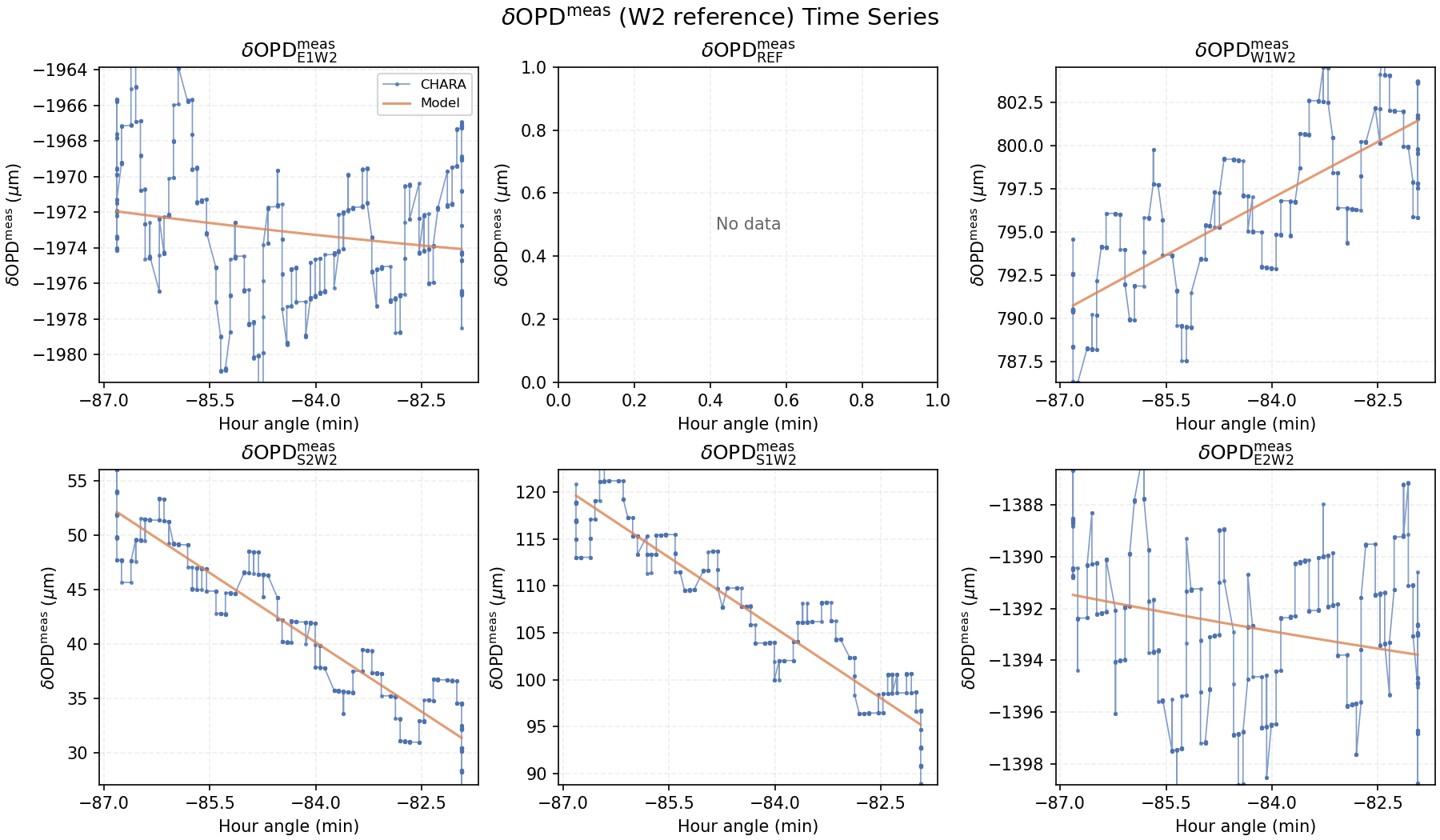}
\caption{Representative dispersion corrected $\delta\mathrm{OPD}^{\rm meas}$ time series for the A--B configuration on
${\rm MJD} = 60918$ with A in MIRC-X and B in MYSTIC (MYSTIC DDLs shown). Blue
points show the measured $\delta\mathrm{OPD}^{\rm meas}$ (W2 reference)
for the six baselines; the orange lines show the geometric model
$\delta\mathrm{OPD}_{\rm geo}(t)=\boldsymbol{B}(t)\cdot(\boldsymbol{r}_A-\boldsymbol{r}_B)$
computed from the time-varying $(u,v)$ and the A--B astrometry. The slow
drift follows the changing projected baseline with hour angle, while
baseline-dependent offsets between data and model indicate residual
H--K dispersion and non-common-path terms. Per-scan scatter is at the
3--4~$\mu$m level, but our astrometric OPD uses time-averaged DDL
positions and fringe tracking residuals, reducing the effective noise to $\sim$0.37~$\mu$m. }
\label{fig:ddl_timeseries_babb_2025Aug31}
\end{figure*}

Figure~\ref{fig:ddl_timeseries_babb_2025Aug31} 
 shows a representative $\delta\mathrm{OPD}^{\rm meas}$ time
series for the A--B configuration (MYSTIC on B, ${\rm MJD} = 60918$). 
Each panel plots the measured
$\delta\mathrm{OPD}^{\rm meas}$ (blue points, W2 reference) for one
baseline and the geometric model
$\delta\mathrm{OPD}_{\rm geo}(t)$
(orange line) computed from the time-varying $(u,v)$ and the adopted
A--B astrometry. 
Over a 5-min window the geometric term can
change by $\sim 20~\mu$m, corresponding to an apparent astrometric shift
of $\Delta \boldsymbol{\theta} \simeq \Delta(\delta\mathrm{OPD})/|\boldsymbol{B}|
\approx 0.02''$ ($\sim 20$~mas along the projected baseline for
$|\boldsymbol{B}|\approx 214$~m), so matching OPD and $(u,v)$ timestamps
is essential. The bootstrap fit of the time series shows that
the mean over each observing window reduces the effective OPD noise to
$\sim 0.37~\mu$m (corresponding to $\sim 0.234$~mas). The A-B/B-A swap solutions
differ by $\sim 175~\mu$as, consistent with the $\lesssim 234~\mu$as uncertainty.

Table~\ref{tab:AB_astrometry_summary} reports five astrometric measurements from five data scans. The unweighted mean formal uncertainty is 0.234\,mas, and the unweighted RMS scatter in $\rho$ about the mean (repeatability) is 0.216\,mas.

\subsubsection{Uncertainty Budget}

For small angles the scalar separation
$\theta = |\boldsymbol{\theta}|$ satisfies
\begin{equation}
    \theta \simeq \frac{\delta\mathrm{OPD}}{|\boldsymbol{B}|},
\end{equation}
so perturbations in baseline length and differential delay contribute to
the 1-$\sigma$ error in quadrature as
\begin{equation}
    \sigma_\theta \simeq
    \left[\left(\theta\,\frac{\Delta|\boldsymbol{B}|}{|\boldsymbol{B}|}\right)^2
    + \left(\frac{\Delta(\delta\mathrm{OPD})}{|\boldsymbol{B}|}\right)^2\right]^{1/2}.
\end{equation}

Longer CHARA baselines reduce the astrometric term as
$\sigma_\theta \propto |\boldsymbol{B}|^{-1}$, providing a geometric
advantage over shorter-baseline interferometric facilities for otherwise comparable
error terms.

For CHARA baselines the geometric term caused by $\Delta|\boldsymbol{B}|$ is small compared to other terms.  
For example, at $\theta = 1.85''$ a typical $10$~mm error observed in baseline length
($\Delta|\boldsymbol{B}|/|\boldsymbol{B}| \sim 3\times10^{-5}$ on
331~m) introduces only $\sim 56~\mu$as of astrometric error.  
By contrast, $\Delta(\delta\mathrm{OPD})\sim250$~nm gives to
$\sim 156~\mu$as at the same baseline.  
We verified lateral pupil shift stability using laboratory AO Shack--Hartmann data \citep[e.g.,][]{Anugu2020b}, and
the pupil images remain stable to within $\sim 1\%$ over the observing
sequence.

Table~\ref{tab:astrometry_error_budget} summarizes the dominant
contributions for the $\alpha$~Psc A--B measurement.  
The quadrature sum yields a total differential-astrometric uncertainty
of $\sim 227~\mu$as at a separation of $1.85''$. 
CHARA does not currently include an end-to-end laser metrology system
analogous to GRAVITY, so non-common-path offsets are not measured
directly and must instead be bounded empirically.

In future, the dispersion systematics could be reduced by using phase tracking available with SPICA-FT \citep{Pannetier2022} and enabling the
CHARA Longitudinal Dispersion Corrector (LDC) hardware correction \citep{Pannetier2021}, increasing the cadence of A-B/B-A and
FT/Science swaps to track time-variable chromatic terms,
along with upgraded DDL hardware to improve repeatability and speed.

While this A-B astrometric precision is modest compared to the best achieved
$\sim25~\mu$as achieved by VLTI/GRAVITY (GRAVITY Collaboration et al. \citeyear{GRAVITY2018A&A...618L..10G}), the
CHARA dual-field mode provides access to northern targets and
separations up to $\sim5''$, offering complementary astrometric reach.

\subsection{Impact of Ba--Bb on the A--B astrometry}\label{app:impactof_BaBb_A_B}

The $\sim 0.23$~mas precision achieved here sets the scale for what
dual-field measurements can reveal about the inner Ba--Bb pair versus
the wide A--B orbit.  
The Ba--Bb separation is only $\sim 7$~mas; on a 331~m baseline this
corresponds to a geometric differential delay of
$\delta\mathrm{OPD}\approx 11~\mu$m.  
For a modest Ba--Bb flux asymmetry ($\lesssim 1\%$) the resulting
photocenter shift is $<0.1~\mu$m, i.e.\ $\lesssim 68~\mu$as, well below
our current $\sim 0.23$~mas dual-field precision.  
Related astrometric wobble measurements in intermediate-mass binaries are
discussed by \citet{Gardner2021,Gardner2022}. 
Operationally, the A--B delay therefore traces the Ba--Bb barycenter
rather than an individual component.
This is consistent with the combined Ba--Bb fringe packets seen in the
science-channel OPD waterfall plots (Figure~\ref{fig:opd_trend}).

In principle, repeated dual-field astrometry over many Ba--Bb orbits
could detect the photocenter wobble of the inner pair, but the expected
semi-amplitude ($A_{\rm wob}\lesssim 0.1$--$0.2$~mas for near-twin
components) is comparable to the present error floor.  
With only two commissioning epochs the Ba--Bb signal is degenerate with
differential-delay systematics and cannot be isolated from the long
A--B motion.  
In practice, the inner orbit is therefore constrained almost entirely by
the resolved Ba--Bb interferometry and archival radial velocities, while
the dual-field A--B astrometry serves primarily to validate the new mode
and confirm that the wide orbit is consistent with the ORB6 solution.

\section{Observing Setup and Data Reduction}
\label{app:obs}

The $\alpha$~Psc system was observed on UT~2025~August~31 (${\rm MJD} = 60918$) and September~01 (${\rm MJD} = 60919$) using all
six CHARA 1-m telescopes \citep{tenBrummelaar2005}.
We employed the standard low-resolution modes of each combiner,
$R_{\rm spec}=50$ for MIRC-X ($H$ band) and $R_{\rm spec}=100$ for
MYSTIC ($K$ band).

\paragraph{Target Acquisition and Fiber Injection}

At each CHARA telescope, the $\alpha$~Psc A--B field is first acquired on the
acquisition and wavefront sensor cameras; then both the TT and AO loops are closed on the bright A component to ensure maximum wavefront stability. In dual-field mode the
telescope delivers the full $\sim5''$ field to the laboratory, so both stars
propagate through the vacuum beam relay and the main delay lines without any
on-sky separation applied upstream (see~Figure~\ref{fig:dual_star_optical_layout}).

Upon arrival in the beam-combiner lab, the pair is imaged by the STST acquisition camera. The field is positioned such that
the midpoint of the A--B pair lies near the optical axis of STST, ensuring that
neither component suffers vignetting from relay optics, dichroics, or pupil
stops.  

The actual separation of A and B into different interferometric channels occurs
only on the MIRC-X + MYSTIC optical table (see~Figure~\ref{fig:dual_star_optical_layout}).  
The MIRC-X instrument is equipped with a 3D translation stage carrying the entrance
single-mode fiber for X, Y, and focus positions. MYSTIC is equipped with a fast steering mirror that helps inject the selected target within the field into the single-mode fiber.   
Using the STST-derived astrometric coordinates, we move the MIRC-X fiber to the photocenter of component~A (the
fringe-tracking star), while the MYSTIC fiber is positioned on
component~B (the science target).  
Fine adjustments are performed with fiber-map scans to maximize
coupling efficiency into both $H$-band (MIRC-X) and $K$-band (MYSTIC)
single-mode fibers.

Once the fibers are centered, the main CHARA delay lines compensate geometric
path differences for the A--B field as a whole, while the internal differential
delay lines specific to MIRC-X and MYSTIC remove the residual $\delta\mathrm{OPD}$
arising from the angular separation between the two stars.  
The resulting configuration provides stable fringe tracking on $\alpha$~Psc~A
and, coherently integrated science fringes on component~B,
enabling both the wide A--B astrometric measurement and the resolved detection
of the Ba--Bb subsystem.

\paragraph{Co-phasing and Differential Delay Alignment}
CHARA does not have a dedicated metrology link connecting the fringe-tracking and science beam combiners and their science interferograms to measure the internal path length between them accurately at the nm level. We co-phase the MIRC-X and MYSTIC instruments with the common light from the Six Telescope
Simulator (STS) in the laboratory; once co-phased, the DDL positions of MIRC-X and MYSTIC were set to zero. For on-sky dual-field mode, the internal DDLs of
the science combiner---the channel receiving $\alpha$~Psc~B---were offset by
the expected geometric delay (Eq.~\ref{eq:dualfield_opd})
corresponding to the A--B separation measured by the STST camera.
We wrote a Python script that takes the separation, position angle, and
CHARA $(u,v)$ coverage to compute the expected OPD/DDL positions.
This ensured that light from both A (fringe tracker) and B (science) arrived within the
coherence window at the start of each scan.

\paragraph{Visibility Calibration}
A single calibrator, HD~13546, with a uniform-disk diameter
$\theta_{\rm UD} = 0.6733 \pm 0.0611~\mathrm{mas}$,
was used for all observations.  
Model checks using \texttt{PMOIRED} confirmed that HD~13546 exhibits no
closure-phase signature or visibility anomalies indicative of binarity.
The same calibrator was observed repeatedly before and after each A--B
science pointing to ensure consistent transfer-function tracking.

\subsection{Data Reduction Pipeline}

\begin{figure*}
\centering
\includegraphics[width=\textwidth]{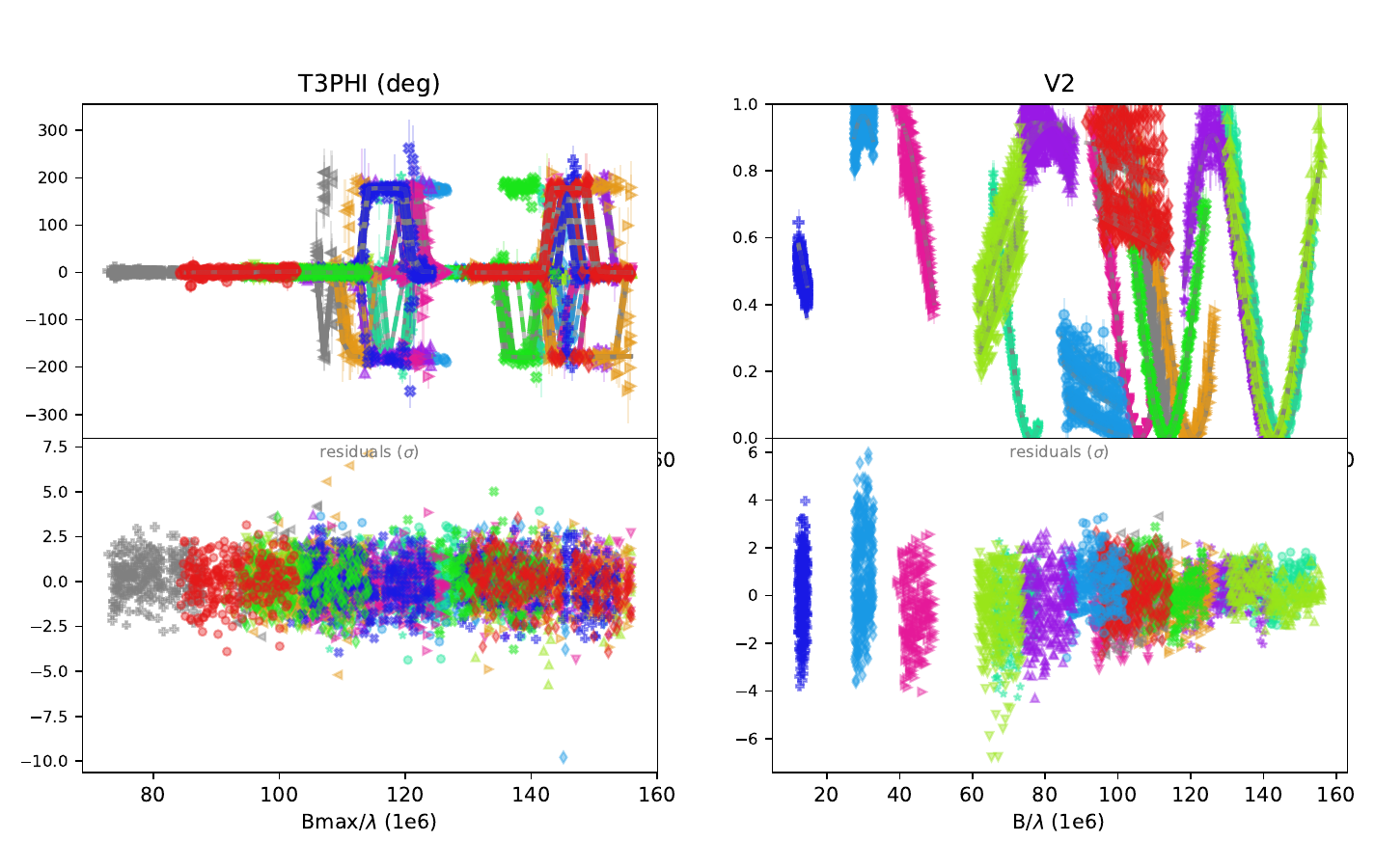}
\caption{PMOIRED binary model fitting residuals for the first MYSTIC epoch (MJD$=60918$). \textbf{Left:} Closure phase data (T3PHI, colored points) and model fit (dashed gray lines) in the upper panel, and the  residuals ($\sigma$) in the lower panel. \textbf{Right:} Same for the visibility-squared data (V2) and model fit. `Bmax' is the maximum projected baseline length for each telescope triangle, B is the projected baseline length for each pair of telescopes, and $\lambda$ is the observed wavelength. 
}
\label{fig:mystic_residuals}
\end{figure*}

Raw interferometric data\footnote{\url{https://www.chara.gsu.edu/observers/database}} were processed using the standard MIRC-X reduction pipelines \citep{Anugu2020}\footnote{\url{https://gitlab.chara.gsu.edu/lebouquj/mircx\_pipeline, v1.5.0}} to obtain uncalibrated,  15 squared visibilities ($V^2$) and 20 closure phases ($T3PHI$), of which 10 are independent. 
Because $\alpha$~Psc is extremely bright, the MYSTIC science channel
saturates before atmospheric decoherence becomes limiting.  The data
presented here were reduced using 10~ms coherent integrations. 
Figure~\ref{fig:opd_trend} shows fringe tracking residuals in the waterfall plots. 
Nonetheless, the stabilized piston delivered by MIRC-X enables coherent
averaging without loss of fringe contrast, producing the clean Ba--Bb
visibility oscillations shown in Figure~\ref{fig:mystic_residuals}.
We reduced datasets with 4~ms integrations for MIRC-X.
Instrumental and atmospheric transfer functions were derived from the calibrator scans using the IDL-based CHARA calibration routines, and applied uniformly to all H- and K-band data. We used a calibration script\footnote{\url{https://www.chara.gsu.edu/tutorials/mirc-data-reduction}} written in IDL by J. D. Monnier to average the measurements over 2.5 minute intervals. The calibrated OIFITS files are available through the JMMC Optical Interferometry Database\footnote{\url{https://oidb.jmmc.fr/index.html}}.

\begin{figure*}
\centering
\includegraphics[width=\textwidth]{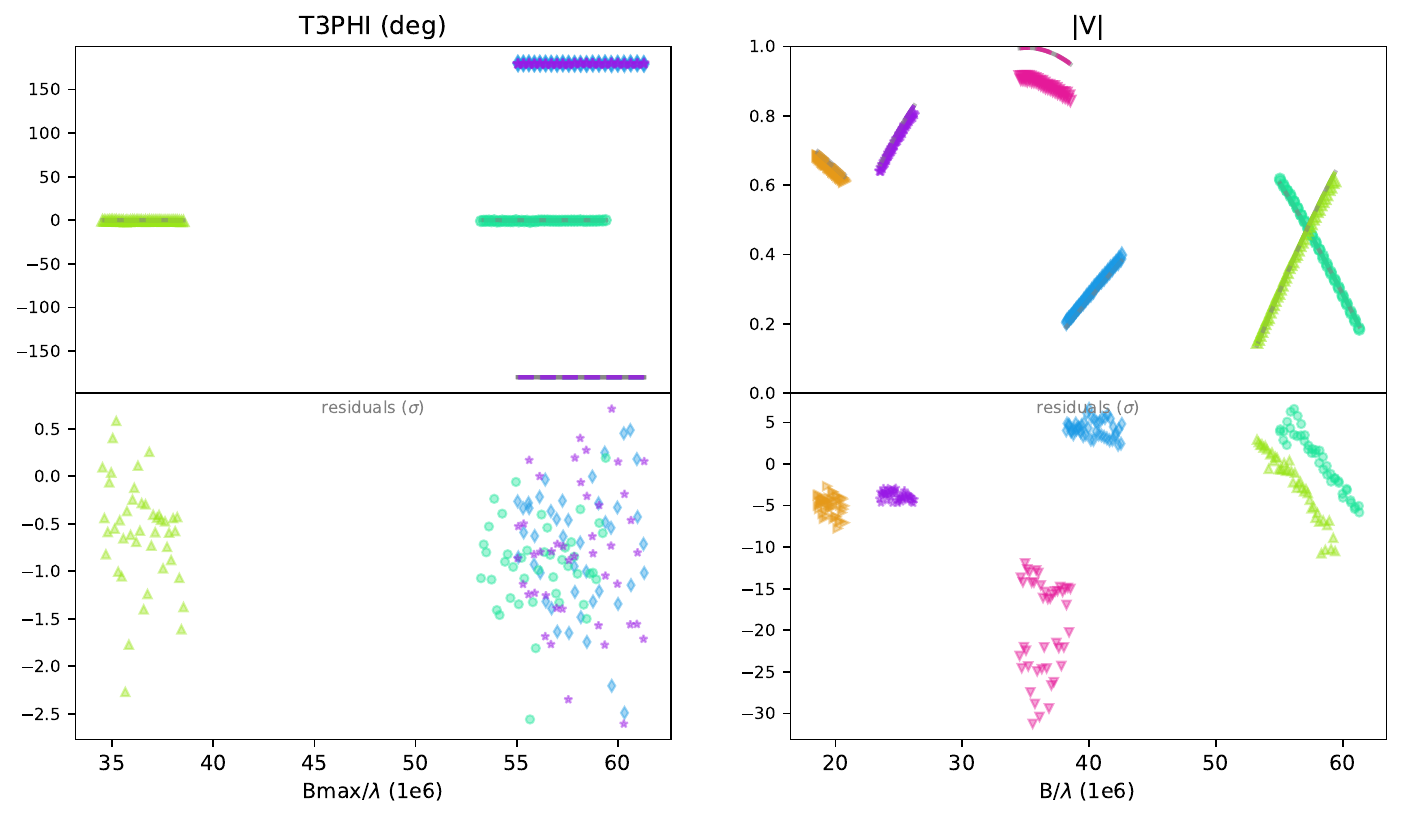}
\caption{Same as Fig.~\ref{fig:mystic_residuals}, but for the GRAVITY epoch ${\rm MJD}=58792$, and displaying visibility ($\vert$V$\vert$) rather than visibility-squared. The large visibility residuals ($\sigma$) are caused by imperfect calibration due to no dedicated calibrator observations being available. }
\label{fig:Gravity_residuals}
\end{figure*}

\section{PMOIRED BINARY FITS FOR $\alpha$ Psc B}\label{app:pmoired}

We conducted a systematic companion search around component~B using \texttt{PMOIRED}\footnote{\url{https://github.com/amerand/PMOIRED.git, v1.4.0}}\citep{Merand2022SPIE12183E..1NM}. 

The free model parameters were the relative astrometric offset of the two components, given by the angular separation ($\rho$) and the position angle (PA, measured from north to east), and the flux ratio $f_{\rm Bb}/f_{\rm Ba}$, where $f$ represent the fluxes of the respective components. In addition, we fitted UD angular diameters of the two components when possible; in practice, this was hindered by the data quality and by the small, barely resolved components. An example of the data, binary fit, and the fit residuals is shown in Figure~\ref{fig:mystic_residuals} for the MYSTIC data from the first epoch.  

The fit to the first MIRC-X and MYSTIC epochs resulted in averaged angular diameters of $\Theta_{\rm UD, Ba} = 0.34\pm0.08$\,mas and $\Theta_{\rm UD, Bb} = 0.30\pm0.08$\,mas. For the second CHARA epoch, which had lower quality data, we were forced to fix $\Theta_{\rm UD, Ba}$ at 0.34\,mas, while fitting only $\Theta_{\rm UD, Bb}$ in the case of MIRC-X data. For the MYSTIC data, we could not fit either diameter and therefore kept both fixed at 0.34 and 0.30\,mas, respectively. As for the GRAVITY data - which, in addition to having lower angular resolution, also lack dedicated calibrators - we could not fit the diameters and kept them fixed at the same values of 0.34 and 0.30\,mas, respectively.

The derived angular diameters correspond to physical stellar radii of $1.76\pm0.41$, and $1.55\pm0.41$, respectively, for Ba and Bb. The measured radii are compatible with the the masses inferred from our orbital solution (Table~\ref{tab:BaBb_orbit_main}); using the main-sequence mass--radius relations compiled by E.\ Mamajek\footnote{\url{http://www.pas.rochester.edu/~emamajek/EEM\_dwarf\_UBVIJHK\_colors\_Teff.txt}} \citep{2013ApJS..208....9P} leads to radii of $\sim1.74 R_\odot$ to $\sim1.73 R_\odot$ for Ba and Bb, respectively, by interpolating in the 
(mass, radius) entries between A9 and F0-V spectral types.

The results from the binary model fit are given in Table~\ref{tab:BaBb_astrometry}. The parameter uncertainties were refined using the bootstrap resampling algorithm implemented in PMOIRED. The fitted angular separations from MIRC-X and MYSTIC were divided by $1.0054$ and $1.0067$, respectively, in order to bring the CHARA instruments onto the same wavelength scale as GRAVITY \citep[][, J. D. Monnier private comm.]{Gardner2022}. For the separation uncertainties derived from the CHARA instruments, we also included an additional $0.2$\% error to account for the uncertainty in wavelength calibration \citep[][, J. D. Monnier private comm.]{Anugu2020,Gardner2022}.

The resulting flux ratios indicate that $\alpha$\,Psc~B is an equal flux binary; averaging the individual measurements results in $f_{\rm Bb}/f_{\rm Ba} = 1.00\pm0.03$. This complicates the astrometric fitting, as there is a $180^{\circ}$ ambiguity in the fitted PA. After visual inspection of the derived positions, we thus had to flip some of the resulting PAs by $180^{\circ}$, so that the bulk of the positions represents realistic orbital motion. The interferometric data are therefore fully compatible with the results in Table~\ref{tab:BaBb_astrometry} being flipped in the PAs (and the flux ratios). The final designation of which component is which rested only on assuming the Ba component has (1) the larger angular diameter in the first MIRC-X and MYSTIC epoch, and (2) the slightly lower velocity semiamplitude in the spectroscopic orbital solution.

\begin{deluxetable}{c c c c c c c c c}
\label{tab:BaBb_astrometry}
\caption{
Relative positions of Bb with respect to Ba measured with GRAVITY, MIRC-X, and MYSTIC.}
\tablehead{
\colhead{MJD} &
\colhead{$\rho$} &
\colhead{PA$^{\rm a}$} &
\colhead{$\sigma_a$} &
\colhead{$\sigma_b$} &
\colhead{PA$_{\rm err}$} &
\colhead{Baselines} &
\colhead{Instrument} &
\colhead{$f_{\rm Bb} / f_{\rm Ba}$} \\
\colhead{} &
\colhead{(mas)} &
\colhead{(deg)} &
\colhead{(mas)} &
\colhead{(mas)} &
\colhead{(deg)} &
\colhead{} &
\colhead{} &
\colhead{}
}
\startdata
60918.471 & 6.864 & 235.780 & 0.014 & 0.014 & 149.5 & E1-E2-S1-S2-W2 & MIRC-X & $1.0198\pm0.0040$\\
60919.369 & 7.068 & 238.041 & 0.014 & 0.014 & 157.1 & E1-E2-S1-S2-W2 & MIRC-X & $0.9754\pm0.0024$\\
60918.431 & 6.903 & 235.494 & 0.014 & 0.014 & 154.0 & E1-E2-S1-S2-W1-W2 & MYSTIC & $0.9859\pm0.0014$\\
60919.388 & 7.139 & 237.824 & 0.014 & 0.014 & 109.3 & E1-E2-S1-S2-W2 & MYSTIC & $0.9993\pm0.0010$\\
58792.153 & 6.434 & 231.397 & 0.018 & 0.003 & 178.7 & A0-G1-J2-K0 & GRAVITY & $0.9979\pm0.0010$\\
58820.095 & 7.443 & 238.077 & 0.102 & 0.054 & 30.6 & A0-G1-J2-J3 & GRAVITY & $0.99^{\rm b}$\\
59183.075 & 2.142 & 117.582 & 0.015 & 0.003 & 179.3 & A0-G1-J2-K0 & GRAVITY & $0.9947\pm0.0010$\\
59834.389 & 2.293 & 157.046 & 0.019 & 0.009 & 12.4 & A0-G1-J2-K0 & GRAVITY & $1.0830\pm0.0502$\\
59889.165 & 4.904 & 220.195 & 0.041 & 0.013 & 4.0 & A0-G1-J2-K0 & GRAVITY & $0.9919\pm0.0027$\\
59894.127 & 6.983 & 237.408 & 0.024 & 0.012 & 17.8 & U1-U2-U3-U4 & GRAVITY & $1.0056\pm0.0053$\\
60648.062 & 7.081 & 246.845 & 0.103 & 0.009 & 164.4 & A0-B2-C1-D0 & GRAVITY & $0.9991\pm0.0026$\\
\enddata
\tablecomments{
PA is measured east of north. $\sigma_a$ and $\sigma_b$ denote the semi-major and semi-minor axes of the astrometric error ellipse, with PA$_{\rm err}$ giving the orientation of the semi-major axis. For the first four CHARA epochs, W1 baselines are omitted because the W1 data were not avilable in those scans.\\
$^{\rm a}$ There is $180^{\circ}$ ambiguity in the PA due to the component having nearly equal fluxes. \\
$^{\rm b}$ The flux ratio fit did not converge for this epoch.
}
\end{deluxetable}

\begin{table}[h!]
\begin{center}
\caption{Revised visual orbit of $\alpha$~Psc A--B.}
\label{tab:AB_orbit}
\begin{tabular}{lcc}
\hline\hline
Parameter & ORB6 & This work \\
\hline
$P_{\rm AB}$ [yr]          & 3270 & $2822 \pm 215$ \\
$a_{\rm AB}$ [arcsec]      & 7.40 & $7.45 \pm 0.35$ \\
$e_{\rm AB}$               & 0.465 & $0.4650 \pm 0.0145$ \\
$i_{\rm AB}$ [deg]         & 113.4 & $113.01 \pm 2.45$ \\
$\omega_{\rm AB}$ [deg]    & 147.9 & $149.72 \pm 5.83$ \\
$\Omega_{\rm AB}$ [deg]    & 3.70 & $3.09 \pm 6.28$ \\
$M^{\rm tot}_{\rm AB}$ [$M_\odot$] & \dots & $5.817 \pm 0.165$ \\
\hline
\end{tabular}
\end{center}
\tablecomments{Columns compare the ORB6 catalogue solution with our new joint fit made with \texttt{orbitize!}\citep{Blunt2020AJ}. Because the wide A--B orbit lacks a radial-velocity constraint on
the node, the visual solution is subject to the usual
$(\omega_{\rm AB},\Omega_{\rm AB}) \leftrightarrow
(\omega_{\rm AB}+180^\circ,\Omega_{\rm AB}+180^\circ)$ degeneracy.  Our fitted
angles are expressed in the \texttt{orbitize!} convention and differ from the
ORB6 values by $\simeq 180^\circ$ rotations and a swap of the argument of
periastron between components, but they represent the same apparent orbit on
the sky.}
\end{table}

\section{Archival Interferometry and Spectroscopy}
\label{app:Archival_data}

\subsection{VLTI/GRAVITY Observations and Data Reduction}

To complement the CHARA measurements, we analyzed seven archival
VLTI/GRAVITY (GRAVITY Collaboration et al. \citeyear{GRAVITY2017}) dual-field observations of the 
$\alpha$~Psc A--B system obtained as part of the VLTI calibration program between 2019 and 2024. In these observations, the A-component was initially used as the phase reference star by the GRAVITY fringe tracker, while the fainter B component was injected 
into the science beam combiner. This was followed by rotation of the field so that the components were swapped, and science beam combiner data were recorded on the A component, while the B component was used as the fringe tracker. The raw data were processed with the official ESO GRAVITY pipeline 
(version~1.7.0; \citealt{Lapeyrere2014}) in the EsoReflex environment \citep{ESOReflex2013}.

Six out of the seven observations were taken with the Auxiliary Telescopes (AT, 1.8\,m diameter), with the remaining one taken with the Unit Telescopes (8\,m diameter). The ATs were used in the `Large' configuration corresponding to a maximum baseline length was $\boldsymbol{B}_{\max} \approx 130$\,m and an angular resolution of $\lambda/2\boldsymbol{B}_{\max} \approx 1.3$\,mas in the K band - in five out the six epochs. The data from MJD$=59894$ were taken in `Small' AT configuration, with $B_{\max} \approx 30$\,m, and $\lambda/2\boldsymbol{B}_{\max} \approx 7.5$\,mas in the K band. The spectral resolution on the science beam combiner was $R\simeq500$ (medium).

The dual-field data lack dedicated calibrator observations, preventing us from obtaining a highly reliable absolute calibration of the visibilities (while the closure phases are little affected). Since A and B components were observed in a sequence, we decided to use the data of component A, which are compatible with an unresolved point source at the VLTI baselines, as the calibrator for component B at each epoch. For the value of the A-component angular diameter used to calibrate the B-component data, we used $0.45\pm0.05$\,mas determined from our first MIRC-X epoch (Section~\ref{sec:A}). While this procedure led to reasonable calibration, which enabled us to use the data for the orbital fit of the Ba-Bb binary, residual systematic errors are present in the visibility data (see Fig.~\ref{fig:Gravity_residuals} for an example).

\subsection{Radial Velocities of the Ba--Bb Subsystem}
\label{app:BaBb_RV}

Radial velocity measurements for the Ba--Bb subsystem were assembled from a
combination of historical spectroscopic observations, archival high-resolution 
spectropolarimetry, and new echelle spectroscopy. The system was first noted 
as a candidate spectroscopic binary by \citet{Frost1929}, and its double-lined 
nature was discovered by \citet{Abt1980ApJS}.  Additional spectroscopic 
measurements by \citet{Abt1985ApJS} confirmed that the system has two narrow-lined 
components of almost identical spectral line appearance and strength.  
Their observations were insufficient to determine an orbit or to assign the 
velocities to specific components. 
We selected the ten doubled-lined measurements
from \citet{Abt1980ApJS} and \citet{Abt1985ApJS} (made between HJD 2440846
and 2444232) for inclusion in our analysis.  These appear with their 
assignment to components Ba and Bb in the beginning of Table~\ref{tab:rv_data}. 

The next set of measurements comes from a collection of spectra hosted at the 
{\it PolarBase}\footnote{https://www.polarbase.ovgso.fr} web site \citep{Petit2014}, 
an archive of data from the TBL/NARVAL high-resolution spectropolarimeter at the 
T\'{e}lescope Bernard Lyot (Pic du Midi Observatory).  We downloaded 56 spectra that
were obtained over 14 nights (between HJD 2456906 and 2456995).  These were re-binned 
onto a standard $\log \lambda$ wavelength grid with a resolving power of 20000.
Significantly, these spectra record two velocity extrema separated by 50 days, 
which was an important clue in determining the orbital period. 

Finally we obtained spectra of $\alpha$~Psc ourselves over the period 2025 Dec 6--8
(HJD 2461015 to 2461017) with the Apache Point Observatory 3.5-m telescope 
and the ARCES instrument.  These observations were made under good seeing conditions
so that spectra of both the A and B components were spatially resolved.  
ARCES covers 3500-10500 \AA\ over 107 echelle orders with a resolving power of 
30000 \citep{Wang2003}.  We reduced the spectra using standard echelle procedures 
in IRAF, removed the blaze function using the procedure in Appendix A of 
\citet{Kolbas2015}, and then merged the orders onto a 1-dimensional $\log \lambda$ grid. 

Radial velocities were measured for the NARVAL and ARCES spectra using a method that
creates cross-correlation functions (CCFs) of each observed spectrum with a model composite 
spectrum over a range in assumed primary velocity and velocity separation 
\citep{Gies1986,Matson2016}.  The model spectrum was adopted from the grid of 
AMBRE high resolution synthetic spectra of \citet{deLaverny2012} for an effective 
temperature of $T_{\rm eff} = 8000$~K, gravity of $\log g = 4.0$, and solar abundances, 
and the spectrum was downloaded from the POLLUX\footnote{https://pollux.oreme.org} web site. 
This model spectrum was adopted for both the Ba and Bb components, and 
a monochromatic flux ratio of 1.0 was assumed for the composite model.  
The wavelength range for the CCF was restricted to a region containing relatively 
isolated and weak metal lines in the range 4474--4578 \AA\ (primarily from transistions 
of \ion{Fe}{1} and \ion{Fe}{2}) in order to create CCFs with minimal broadening.  
The derived velocities and their uncertainties are listed in Table~\ref{tab:rv_data}.
We caution that the NARVAL spectra generally also record the spectral lines of the 
A component, but these are broad enough that they have little impact on the measurements 
of the narrow-lined Ba and Bb components.  We estimate that any blending issues with the 
A component will introduce shifts that are comparable to the uncertainty estimates for the 
individual radial velocities. 

We initially used the orbital fitting code RVFIT \citep{Iglesias-Marzoa2015} to fit 
the radial velocity curves of Ba--Bb and to identify those cases where the 
measurements were swapped between the nearly identical components.   
The fitted radial velocity curves are shown in the lower panel of Figure~\ref{fig:BaBb_orbit}. 
The radial velocity orbits were subsequently combined with the astrometric orbit
(Appendix~\ref{app:BaBborbit}) to arrive at the physical elements reported in Table~\ref{tab:BaBb_orbit_main}.

\section{Orbital fitting methods}\label{app:orbit}

\subsection{Ba--Bb orbit}\label{app:BaBborbit}

We used the orbfit-lib IDL library \footnote{\url{https://www.chara.gsu.edu/analysis-software/orbfit-lib}} \citep[orbfit-lib;][]{2006AJ....132.2618S,2016AJ....152..213S}. to fit all of the three individual solutions in Table~\ref{tab:BaBb_orbit_main}. For the combined astrometric + spectroscopic solution, we scaled the measurement errors so that both the astrometric and RV datasets contribute approximately the same amount to the total residuals.

\subsection{A--B orbit}\label{app:ABorbit}

We modeled the long--period $\alpha$~Psc A--B visual orbit using the 
\texttt{orbitize!} package \citep{Blunt2020AJ}, which performs Bayesian 
inference of Keplerian orbits from relative astrometry.  
The historical measurements were extracted from the Washington Double Star 
Catalog (WDS) and reformatted into the \texttt{orbitize!} input table with 
separation, position angle, and their uncertainties.  
We adopted the combined dynamical mass 
($M_{\rm A}+M_{\rm B}$) and parallax as Gaussian hyperparameters in the fit.
Specifically, we used $M_{\rm tot}=5.86\pm0.12\,M_\odot$ from the A+Ba+Bb
masses in Table~\ref{tab:system_properties} and a distance prior
$d=48.17\pm0.34$~pc from the Ba--Bb dynamical solution, equivalent to
$\varpi=20.76\pm0.15$~mas.

Because the available astrometry spans only $\sim2\%$ of the 
$\sim3300$~yr orbit, we imposed tight Gaussian priors centered on the ORB6 
solution to prevent unphysical degeneracies in
$(\Omega_{\rm AB}, \omega_{\rm AB}, i_{\rm AB}, a_{\rm AB}, e_{\rm AB})$.
We used the affine--invariant MCMC sampler implemented in 
\texttt{orbitize!} with 15,000 total steps and a 5,000--step burn--in.  
Posterior samples were used to draw sky--plane orbits, generate 
separation/PA curves, and construct the corner--plot visualizations.  
The median posterior orbit closely reproduces the ORB6 ellipse and 
matches all modern CCD and interferometric measurements. Table~\ref{tab:AB_orbit} presents the orbital solution.

\begin{figure}
    \centering
\includegraphics[width=0.47\textwidth]{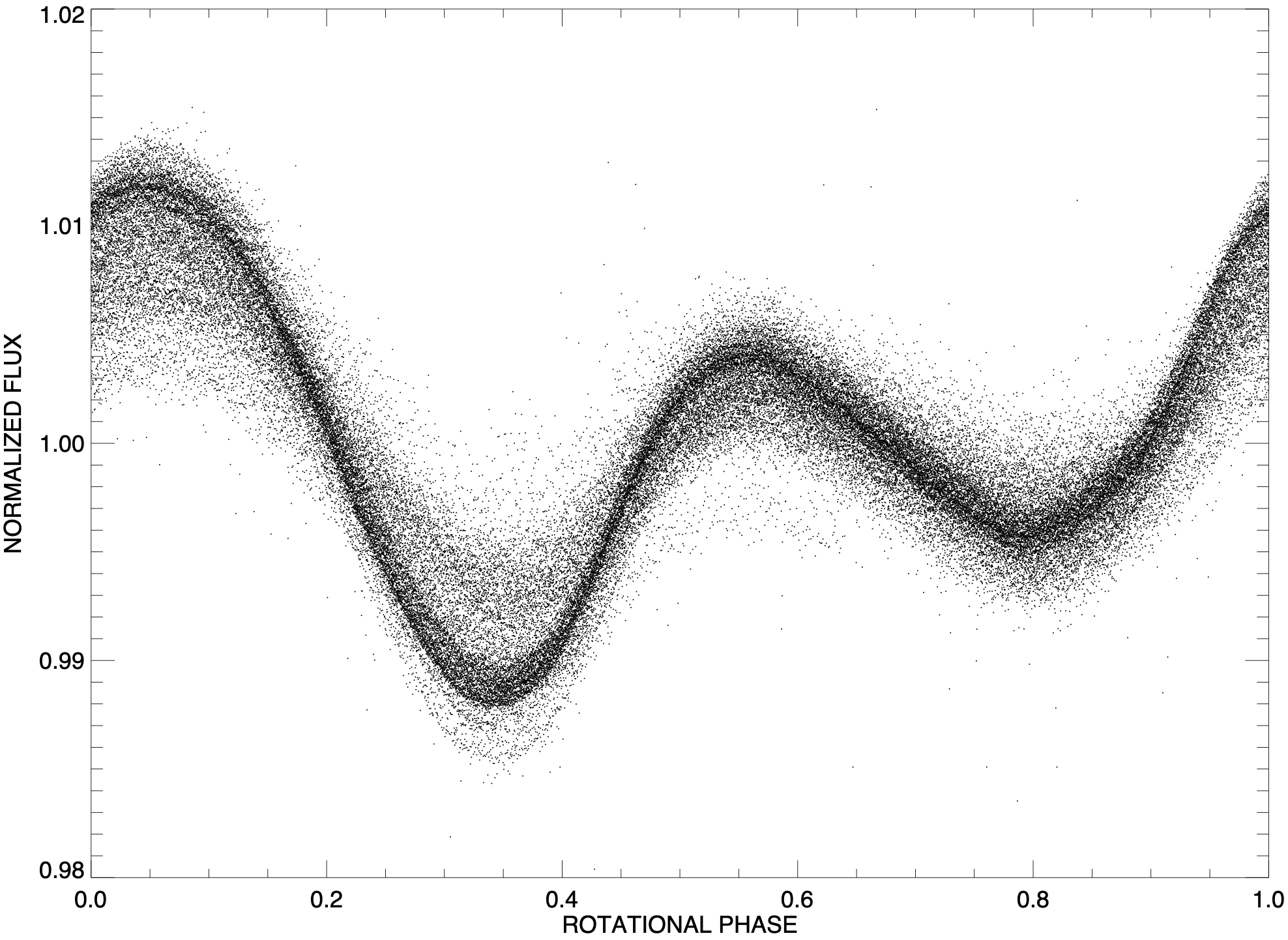}
    \caption{
        Phase-folded TESS light curves for $\alpha$~Psc at $P = 1.490975$\,d. Flux is normalized to unity in each sector by dividing by the median.
    }
    \label{fig:tess_folded}
\end{figure}

\begin{figure}
    \centering    
    \includegraphics[width=0.49\textwidth]{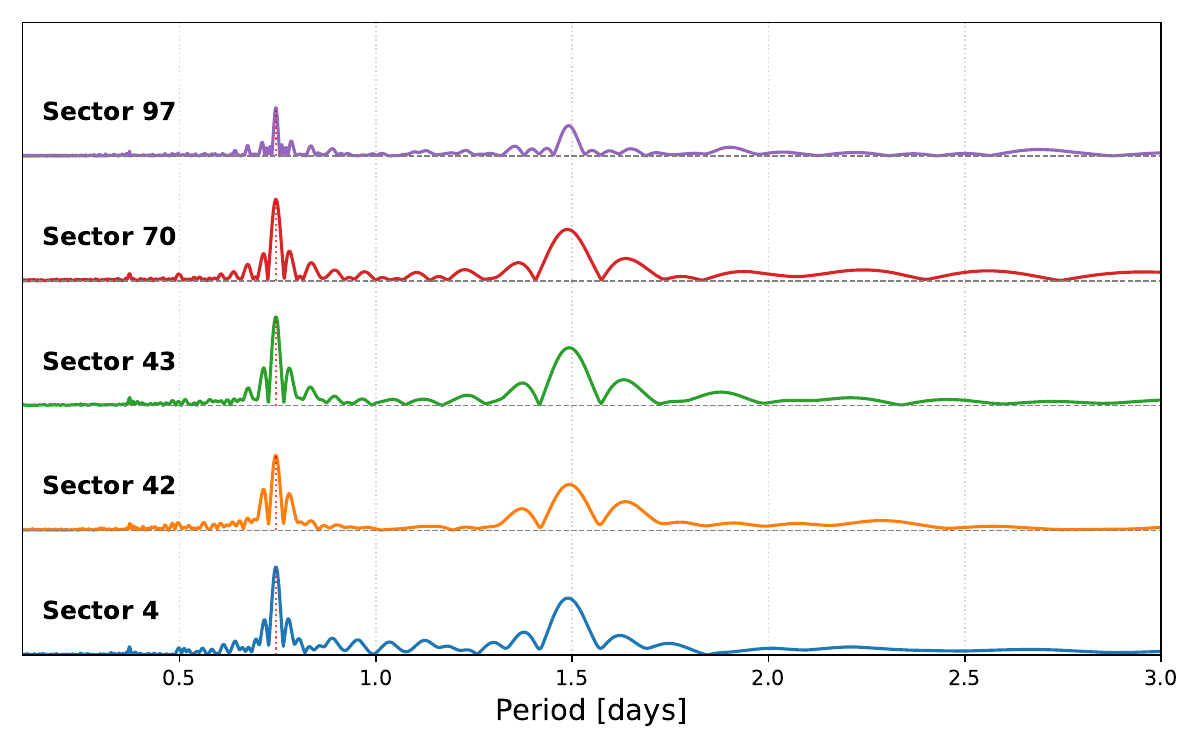}
    \caption{
        Sector-by-sector Lomb--Scargle periodograms for $\alpha$~Psc.
        Each curve (offset vertically for clarity) corresponds to one TESS sector.  
        All five sectors exhibit a dominant peak at $0.7455$\,d, with a secondary peak at $1.490975$\,d.  
        No significant power is detected at longer periods.
    }
    \label{fig:tess_ls}
\end{figure}

\section{TESS Photometric Analysis: $\alpha$~Psc~A}
\label{app:tess}

Here we describe the processing of the five TESS sectors used to characterize the photometric variability of the A component and its relation to the established rotation period $P_{\mathrm{rot}} = 1.490975$~d \citep{Borra1980ApJS...42..421B,Sikora2019MNRAS.483.2300S,Das2022ApJ...925..125D}.

We analyzed all publicly available TESS short-cadence (120\,s) observations of 
$\alpha$~Psc (HD~12447), spanning Sectors~5 (2018), 42--43 (2021), ~70 (2023) and 97 (2025).  
Light curves were obtained via the \texttt{lightkurve} \citep{Lightkurve2018} interface to MAST using the \texttt{SPOC} PDCSAP\_FLUX products.  
For each sector, we performed the following steps:

We downloaded the calibrated PDCSAP\_FLUX time series for each TESS sector, which removes long-term instrumental trends and contamination while preserving intrinsic stellar variability. All cadences flagged by the SPOC quality bitmask were removed, along with residual NaNs, and each sector was median-normalized to place the light curves on a common relative-flux scale. Lomb--Scargle periodograms were then computed using the \texttt{lightkurve.periodogram} interface over 0.1--5\,d with an oversampling factor of five. In every sector, the highest peak occurs at $\simeq 0.7455$\,d, with a secondary peak at $1.490975$\,d. We treat the 0.7455~d signal as the first harmonic and adopt the established rotational period $P_{\mathrm{rot}} = 1.490975$\,d \citep{Borra1980ApJS...42..421B,Sikora2019MNRAS.483.2300S,Das2022ApJ...925..125D} as a working ephemeris.  Finally, the light curves were phase-folded on this period using the epoch of phase zero 
from \citet{Das2022ApJ...925..125D}. 

The resulting summary (Figures~\ref{fig:tess_folded} and ~\ref{fig:tess_ls}) shows a stable double-wave morphology across all five sectors, despite being separated by nearly seven years of TESS operations; the harmonic structure in these figures is consistent with rotational modulation.  
No statistically significant power is detected at longer timescales ($P \gtrsim 3$\,d).  
The stability of the signal across all sectors is consistent with a persistent phenomenon in component~A and is unlikely to be related to the newly resolved Ba--Bb subsystem detected in our CHARA dual-field observations.

The Lomb--Scargle periodograms for all five TESS sectors show a sequence of 
peaks at $P_{\mathrm{rot}}/2$ and $P_{\mathrm{rot}}$, 
a pattern that can be produced by rotational modulation.
The strongest peak occurs at $P_{\mathrm{rot}}/2 \simeq 0.7455$\,d, with a 
significant peak at $P_{\mathrm{rot}} = 1.490975$\,d.  When folded on the full rotation period, the light curves exhibit a persistent double-wave morphology whose relative lobe amplitudes vary between sectors, consistent with evolving surface structure.
The double-wave rotational light curve can arise from 
surface inhomogeneities related to the magnetic field.

Future high-angular-resolution observations in J~band with MIRC-X and  visible-wavelength (R~band) observations with SPICA~\citep{Mourard2017} will provide both higher 
spatial resolution and enable direct imaging tests of the surface  structure implied by the TESS light curves.

\begin{deluxetable*}{lcccc}
\digitalasset
\tablecaption{Radial velocity measurements used in the orbit fitting.
Times are heliocentric Julian dates (HJD, days).
\label{tab:rv_data}}
\tablehead{
\colhead{HJD} &
\colhead{RV$_1$} &
\colhead{$\sigma_{\mathrm{RV}_1}$} &
\colhead{RV$_2$} &
\colhead{$\sigma_{\mathrm{RV}_2}$} \\
\colhead{} &
\colhead{(km\,s$^{-1}$)} &
\colhead{(km\,s$^{-1}$)} &
\colhead{(km\,s$^{-1}$)} &
\colhead{(km\,s$^{-1}$)}
}
\startdata
2440846.755  &  31.40 &  1.40 & -21.60 &  1.40 \\
2443055.827  &  -34.40 &  2.10 & 46.70 &  1.10 \\
2443055.965  &  -38.60 &  2.20 & 59.30 &  1.70 \\
2443459.776  &  -32.40 &  1.60 & 46.80 &  1.80 \\
2443460.722  &  -21.10 &  2.00 & 30.30 &  1.30 \\
2443808.884  &  -46.30 &  1.30 & 58.20 &  1.80 \\
2443809.975  &  -33.00 &  1.40 & 39.40 &  3.00 \\
2443810.853  &  -18.60 &  3.10 & 26.00 &  2.10 \\
2443831.687  &  -77.40 &  1.60 & 87.50 &  2.50 \\
2444232.630  &  -75.10 &  0.90 & 85.50 &  2.40 \\
2456906.597  &  -83.33 &  1.53 & 92.12 &  1.86 \\
2456906.599  &  -83.89 &  1.87 & 92.54 &  2.50 \\
2456906.600  &  -84.20 &  1.81 & 93.41 &  2.50 \\
2456906.602  &  -83.92 &  1.88 & 91.64 &  2.51 \\
2456925.554  &  37.23 &  1.57 & -28.14 &  1.83 \\
2456925.556  &  37.31 &  1.52 & -28.28 &  1.75 \\
2456925.557  &  37.28 &  1.61 & -28.38 &  1.81 \\
\enddata
\tablecomments{This Table  is published in its entirety in the electronic 
version.  A portion is shown here 
for guidance regarding its form and content. }
\end{deluxetable*}

\bibliography{sample701.bib}
\bibliographystyle{aasjournalv7}
\end{document}